\documentclass[11pt,fleqn,a4paper]{article}
\usepackage{graphicx}
\usepackage{fullpage}
\usepackage[authoryear]{natbib}
\usepackage{amssymb}
\usepackage{amsthm}
\usepackage{epstopdf}
\usepackage{amsmath}
\usepackage{ctable}
\usepackage{subfigure}
\usepackage{setspace}
\usepackage[bottom]{footmisc}
\usepackage{endnotes}
\usepackage{graphicx}
\usepackage{textcomp}
\usepackage{natbib}
\usepackage[ansinew]{inputenc}
\usepackage{verbatim}
\usepackage{lscape}
\usepackage{array}
\usepackage{fltpoint}
\usepackage{dcolumn}
\usepackage{booktabs}
\usepackage{rotating}
\usepackage[norounding,point]{rccol}
\usepackage{hyperref}
\usepackage{placeins}
\usepackage{multirow}
\usepackage{paralist}
\usepackage{enumitem}

\usepackage[margin=1in]{geometry}

\numberwithin{equation}{section}

\hypersetup{
colorlinks=true,
urlcolor=blue,
linkcolor=red,
citecolor=blue
}

\usepackage{setspace}
\usepackage{xspace}

\newtheorem{theorem}{Theorem}

\newtheorem{assumption}{Assumption}

\begin{document}

\begin{center}
{\large \vspace{0.4cm}}

{\Large Martin Huber* \& Jannis Kueck+}\medskip

{\small {*University of Fribourg, Department of Economics\\ + Heinrich Heine University D\"usseldorf, D\"usseldorf Institute for Competition Economics 
} \smallskip }
\end{center}

\thispagestyle{empty}
\vspace*{0.2 cm}
\begin{abstract}
\noindent This study demonstrates the existence of a testable condition for the identification of the causal effect of a treatment on an outcome in observational data, which relies on two sets of variables: observed covariates to be controlled for and a suspected instrument. Under a causal structure commonly found in empirical applications, the testable conditional independence of the suspected instrument and the outcome given the treatment and the covariates has two implications. First, the instrument is valid, i.e. it does not directly affect the outcome (other than through the treatment) and is unconfounded conditional on the covariates. Second, the treatment is unconfounded conditional on the covariates such that the treatment effect is identified. We suggest tests of this conditional independence based on machine learning methods that account for covariates in a data-driven way and investigate their asymptotic behavior and finite sample performance in a simulation study. We also apply our testing approach to evaluating the impact of fertility on female labor supply when using the sibling sex ratio of the first two children as supposed instrument, which by and large points to a violation of our testable implication for the moderate set of socio-economic covariates considered.
\end{abstract}

\smallskip\\
{\scriptsize 
The authors are grateful to Joachim Freyberger, Leonard Henckel, Peter Hull, Toru Kitagawa, Christoph Rothe, and conference/seminar participants at the Causal Data Science Meeting (2022), the Annual Meetings of the German Economic Association (2022; Basel), the Conference on Counterfactual Methods for Policy Impact Evaluation (2022; Mannheim), and research seminars at the Luxembourg Institute of Socio-Economic Research (2022) and the University of T\"{u}bingen (2022) for their valuable comments. The authors have no conflict of interest to declare. Addresses for correspondence: Martin Huber, University of Fribourg, Bd.\ de P\'{e}rolles 90, 1700 Fribourg, Switzerland; martin.huber@unifr.ch. Jannis Kueck, Heinrich Heine University D\"usseldorf, Universit\"atsstra\ss e 1,
40225 D\"usseldorf, Germany; kueck@dice.hhu.de.}

\newpage
\setcounter{page}{1}
\section{Introduction}

Causal inference methods for assessing the effect of a treatment or policy intervention (e.g.\ a training or a marketing campaign) on an outcome variable (e.g.\ earnings or sales) typically make use of identifying assumptions that are deemed untestable. For instance, the popular selection-on-observables, unconfoundedness, conditional independence, or ignorability assumption imposes that the treatment is as good as randomly assigned after controlling for observed covariates, see e.g.\ \citet{ImWo08} for a review of evaluation approaches in this context. Whether the set of covariates is sufficient for this assumption to hold is conventionally motivated based on theory, intuition, domain knowledge, or previous empirical findings.

In this paper, we demonstrate the existence of a testable condition for the satisfaction of the selection-on-observables assumption in observational data in empirically relevant causal models, which implies that identification can be tested in the data. This condition relies on two types of observed characteristics, namely  covariates to be controlled for and a so-called \textit{suspected} instrument: a variable that can potentially serve as an instrument, even though it is \textit{a priori} unknown whether it satisfies specific instrument validity assumptions. The testable condition arises if the following assumptions hold: First, there is no reverse causality from the outcome to the treatment, covariates, or the \textit{suspected} instrument and from the treatment to the covariates or the \textit{suspected} instrument. Second, the \textit{suspected} instrument is statistically associated with the treatment conditional on the covariates, e.g.\ through a first stage effect of the \textit{suspected} instrument on the treatment. If the \textit{suspected} instrument is conditionally independent of the outcome given the treatment and the covariates in this context, then the instrument and the treatment satisfy the following two conditions: (A) the instrument is valid, i.e.\ does not directly affect the outcome (other than through the treatment) and is not associated with unobservables affecting the outcome conditional on the covariates and (B) the selection-on-observables assumption holds w.r.t. the treatment, i.e.\ the treatment is not associated with unobservables affecting the outcome conditional on the covariates. Therefore, the conditional independence of the \textit{suspected} instrument implies the identification of treatment effects based on the selection-on-observables assumption (B). When focusing on average causal effects  like the average treatment effects (ATE), it is sufficient to test mean (rather than full) conditional independence of the outcome and the \textit{suspected} instrument.

It is worth noting that the jointly tested instrument validity (A) does not imply the identification of causal effects without further assumptions like monotonicity of the treatment in the instrument, see \cite{Imbens+94}. For this reason, the \textit{suspected} instrument only serves as an auxiliary variable to test identification based on the selection-on-observables assumption (A), but we do not use it for instrument-based identification, e.g., of the local average treatment effect (LATE). If, however, the researcher is convinced to have an instrument satisfying all identifying assumptions, including (A), our approach serves as overidentification test for the additional satisfaction  of the selection-on-observables assumption (B). Basing effect estimation on (B) rather than (A) can be beneficial in terms of precision (as instrument-based methods are generally less efficient) and external validity (as the ATE refers to the total population while the LATE only refers to a subpopulation). 

There is an extensive literature on conditional independence tests of variables. Closely related to our testing problem, \cite{fan1996consistent}, \cite{racine1997consistent} and \cite{racine2006testing} provide methods to test the significance of regressors using kernel methods in semi- and nonparametric regression models.  Further specification tests have been suggested in \cite{bierens1982consistent}, \cite{hardle1993comparing}, \cite{horowitz1994testing} and \cite{wooldridge1992test}. All of these methods assume the regressors to be fixed (i.e.\ \textit{a priori} chosen by the researcher), while the testing approaches suggested in this paper permit selecting important control variables in a data-driven way. More specifically, we construct tests for the conditional independence of the instrument using doubly robust (DR) methods, see \citet{Robins+94} and \citet{RoRo95}. In our context, these DR approaches are based on statistical models for both the instrument and the outcome. We apply the double machine learning (DML) framework of \cite{doubleML}, in which the models for the instrument and the outcome are learned in a data-driven way based on machine learning. This appears particularly attractive in high-dimensional settings where the number of available covariates is relatively large and/or when the instrument and outcome models include high-dimensional interactions and polynomial functions of the covariates to allow for nonlinear associations. 

Under specific regularity conditions, we show that our testing approach is root-n-consistent despite the use of flexible machine learning methods.  Following \citet{AtheyImbens2016}, \citet{WagerAthey2018}, \citet{AtheyTibshiraniWager2019}, and \citet{leeetal2020}, we also suggest a machine learning-based algorithm for detecting heterogeneity in the violations of the conditional independence as a function of the covariates. This permits detecting subgroups in which the violations are particularly large, in order to increase (asymptotic) testing power. Finally, we propose a test based on the average squared violation across all covariate values, which represents a global testing approach to account for covariate-dependent heterogeneity in violations. We show that under the null hypothesis, this test also satisfies so-called \citet{Neyman1959}-orthogonality, implying that we may account for covariates by machine learning without compromising on a desirable asymptotic behavior, given that specific regularity conditions hold. 

In a simulation study with 50 covariates, we find that the various tests perform decently in terms of empirical size and power, even under moderate sample sizes of 1000 or 4000 observations.
As an empirical illustration, we apply our testing approach to US census data previously considered in \citet{Angrist+98} for assessing the impact of fertility, which is the treatment, on female labor supply  when using the sex ratio of the first two children as instrument. The intuition for this instrumental variable (IV) strategy is that if parents tend to have a preference for mixed sex children, then having two children of the same sex, which is arguably randomly assigned by nature, increases the chances of getting a third child. Based on findings in  \citet{RoWo00} and \citet{Lee07f}, one might, however, challenge whether all identifying assumptions required for the IV-based identification of causal effects are satisfied. Here, we do not impose any IV assumptions \textit{a priori}, but use the sibling sex ratio to test the joint satisfaction of IV validity and the selection-on-observables assumption. Our results point to a violation of the conditional independence of the instrument under our moderate set of covariates, which consists of several socio-economic characteristics like mother's age and father's income. Therefore, testing suggests that the treatment does not satisfy the selection-on-observables assumption, or the sex ratio is not a conditionally valid instrument, or both.

This paper connects to several strands of the causal inference literature. Most closely related are studies assuming an instrument that is conditionally valid given covariates, in order to test the selection-on-observables assumption on the treatment based on the same or related conditions as in this paper, see \citet{deLunaJohansson2012}, \citet{BlackJooLaLondeSmithTaylor2015}, \citet{chen2018testing}, and \citet{feve2018estimation}. Here, we highlight that (A) and (B) can be tested jointly, such that one need not impose the existence of a valid instrument prior to testing. In contrast to previous work, we establish that under specific assumptions, the testable condition is both necessary and sufficient for the joint satisfaction of conditions (A) and (B). Second, unlike the testing methods in the prior literature, we propose DR testing approaches based on DML that accommodates high-dimensional covariates. We note that a joint test is also proposed by \citet{angrist2017leveraging} in the context of linear regression models, whereas this study suggests nonparametric testing methods. Furthermore, our approach relates to \citet{angrist2015wanna}, who consider the opposite scenario of a treatment satisfying a  selection-on-observables assumption (B) in order to test whether a further variable is a valid instrument (A). In the absence of a selection-on-observables assumption on the treatment, specific instrument assumptions are only partially testable; see, for instance the methods proposed by \citet{Kitagawa2008}, \citet{HuMe11},  \citet{MoWa2014}, and \citet{Guber2018}.

Relatedly, one strand of the statistics literature imposes the selection-on-observables assumption (B) and exploits the conditional independence condition to identify subsets of covariates that are sufficient for identification, implying that the remaining covariates satisfy instrument validity (A); see, for instance, \citet{de2011covariate} and \citet{vanderweele2011new}. Again, our approach differs in that it tests the selection-on-observables and instrument validity assumptions jointly, rather than testing the latter conditional on the former. Closer to our framework, \citet{entner2013data} do not pre-impose the selection-on-observables assumption when searching for a sufficient set of covariates to control for based on conditional independence and consider a parametric modeling approach for testing, however, without providing asymptotic guarantees. Here, we provide machine learning approaches that flexibly accommodates nonparametric models and potentially high-dimensional covariates and have desirable asymptotic properties under certain regularity conditions.

Our method of testing identification addresses in some sense a statistical problem  `in between' classical treatment evaluation, where both the treatment and the identifying assumptions are predetermined, and causal discovery, see e.g.\ \citet{KalischB2014}, \citet{peters2017elements}, \citet{Glymouretal2019}, or \citet{breunig2021testability}. Causal discovery typically does not pre-define the treatment and outcome, but aims at learning the causal relations between two or more variables in a data-driven way, possibly under parametric restrictions or the assumption that all relevant variables in the causal system (apart from random error terms) are observed. Here, we do not rely on such assumptions, but instead impose more causal structure to distinguish the treatment, outcome, covariates, and the \textit{suspected} instrument. This structure appears realistic in many empirical contexts with information about the timing of variable measurement. For instance, a treatment taking place in an earlier period can affect and outcome measured in a later period, but not vice versa. In contrast to classical treatment evaluation, we do, however, not pre-impose specific identifying assumptions, but test them in the data.

The remainder of this study is organized as follows. Section \ref{Assumptions} discusses a set of identifying assumptions which are used to derive a testable implication of the joint satisfaction of IV validity and selection-on-observables assumptions implying the identifiability of treatment effects. Section  \ref{meanind} provides a modified testable implication when assuming that the IV validity and selection-on-observables assumptions hold with respect to the mean potential outcomes, rather than the entire potential outcome distributions. 
Section \ref{DRtest} proposes several approaches to test identification. First, a test based on DML is considered to control for (possibly high-dimensional) covariates in a data-driven way. DML permits verifying whether the testable implication holds on average in the total sample or within subsamples defined as a function of the covariates. Second, a test based on average squared violations, i.e.\ squared differences in conditional mean outcomes, is proposed which jointly tests for violations of the testable implication across all covariate values. Then, Section \ref{simulation} provides a simulation study. The application to the evaluation of the impact of fertility on female labor supply is given in Section \ref{application}. 
Section \ref{co} concludes.


\section{Assumptions and Testable Implication}\label{Assumptions}

Of interest is the causal effect of a treatment $D$ on an outcome $Y$, and both variables might be discretely or continuously distributed. In our discussion of causality, we will make use of the potential outcome framework as for instance advocated in \cite{Neyman23} and \cite{Rubin74}. We denote by $Y(d)$ the potential outcome when exogenously setting the treatment $D$  of a subject to some value $d$ in the support of the treatment. More generally, we will use capital and lower case letters for referring to random variables and specific values thereof, respectively. Importantly, denoting the potential outcome $Y(d)$ as a function of a subject's treatment status $D=d$ alone implicitly imposes that (i) someone's potential outcomes are not affected by the treatment status of others and (ii) there are no different versions of any treatment level $d$ across individuals. This is known as the `Stable Unit Treatment Value Assumption' (SUTVA), see for instance the discussion in \citet{Rubin80} and \citet{Cox58}, and is imposed throughout this paper. Furthermore, we denote by $X$ a vector of observed covariates to be used as control variables and by $Z$ one or several observed instrumental variables, whose properties are yet to be defined. Finally, let $\mathcal{X}$, $\mathcal{Z}$, $\mathcal{D}$, and $\mathcal{Y}$ denote the support of $X$, $Z$, $D$, and $Y$, respectively. We will subsequently discuss the assumptions which permit testing identification.

Our first assumption imposes some structure concerning which variables may causally affect other variables. It also states that any two variables which are associated with each other via causal paths (possibly conditional on other variables) are necessarily statistically dependent, which is known as causal faithfulness and formally defined below.

\begin{assumption}[Causal structure and faithfulness]\label{ass0}
\begin{eqnarray*}
&&D(y)=D,\quad X(d,y)=X,\textrm{ and }Z(d,y)=Z\quad  \forall d \in \mathcal{D}\textrm{ and }y  \in \mathcal{Y}.
\end{eqnarray*}
Only variables which are d-separated in some causal model are statistically independent.
\end{assumption}

\begin{figure}[!htp]
	\centering \caption{\label{causalchain}  Causal graph satisfying Assumption \ref{ass0}}\bigskip
\centering \includegraphics[width=1\textwidth]{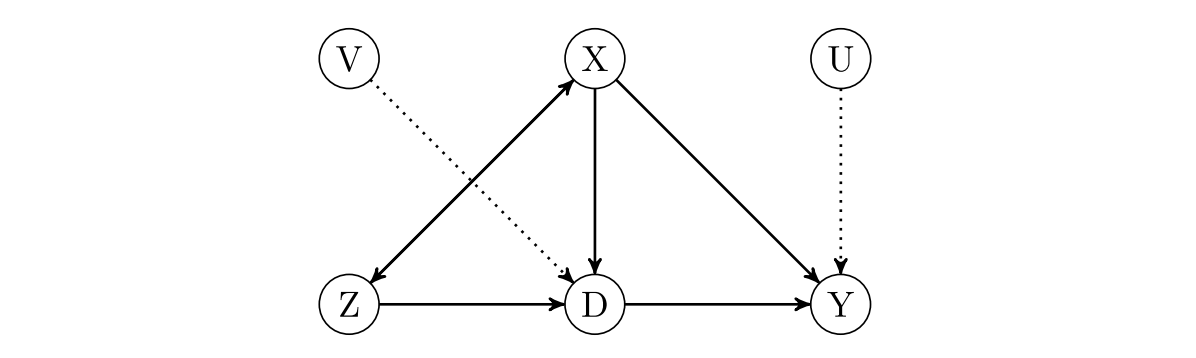}	
\end{figure}

\noindent The first line of Assumption \ref{ass0} rules out reverse causal effects of outcome $Y$ on $D$, $X$, or $Z$. For example, this means that the treatment $D$ is not causally affected by the outcome variable: $D(y)=D$. It also states that the treatment $D$ must not causally affect $X$ or $Z$, while both $X$ and $Z$ might affect $D$, $Y$, or each other. These conditions are satisfied in the causal framework in Figure \ref{causalchain}, which represents the causal associations between instrument $Z$, treatment $D$, and outcome $Y$ by arrows in a directed acyclic graph (DAG), as e.g.\ considered in \citet{Pearl00}. In the DAG, $Z$ and $X$ affect $D$ and $D$ and $X$ affect $Y$. Furthermore, $X$ might affect $Z$ or vice versa, as indicated by the bidirectional causal arrow. This scenario is in line with the practice of measuring covariates and instruments prior to treatment assignment, which rules out reverse causality of $D$ and $Y$ on the pre-treatment $X$ and $Z$. The DAG also includes the unobserved terms $U$ and $V$ which affect $Y$ and $D$, respectively, with the dashed arrows indicating that these effect cannot be observed.

The second line of Assumption \ref{ass0} imposes causal faithfulness, meaning that only variables that are d-separated, i.e.\ not connected by any causal paths (possibly conditional on other variables), are statistically independent (or conditionally independent). The d-separation criterion is a general rule for determining in nonparametric models whether two variables are causally unrelated, possibly conditional on third variables. In causal analysis, this enables judging the nonparametric identifiability of causal effects under specific conditioning sets. For example, in an observational study, if the treatment and the potential outcomes are d-separated given certain covariates, it implies that conditioning on the covariates appropriately controls for confounding, ensuring that the treatment and potential outcomes are conditionally independent. Thus, in nonparametric causal models, d-separation helps identify which variables need to be controlled for to estimate causal relationships correctly.

More formally, the d-separation criterion of \citet{pearl1988probabilistic} relies on blocking causal paths between variables (represented by arrows) to render them independent. A path between two (sets of) variables $A$ and $B$ is blocked when conditioning on a (set of) control variable(s) $C$ if 
\begin{enumerate}
\item the path between $A$ and $B$ is a causal chain, implying that $A\rightarrow M \rightarrow B$ or $A\leftarrow M \leftarrow B$, or a confounding association, implying that $A\leftarrow M \rightarrow B$, and variable (set) $M$ is among the control variables $C$ (i.e.\ controlled for),
\item the path between $A$ and $B$ contains a collider, implying that $A\rightarrow  S  \leftarrow B$, and variable (set) $S$ or any variable (set) causally affected by $S$ is not among control variables $C$ (i.e.\ not controlled for).
\end{enumerate}
Based on this definition of blocking, the d-separation criterion states that $A$ and $B$ are d-separated (i.e.\ causally independent or unrelated) when conditioning on control variable(s) $C$ if and only if $C$ blocks all paths between $A$ and $B$. d-separation is sufficient for the (conditional) statistical independence of two variables, which will be useful to prove Theorems \ref{theorem1} and \ref{theorem2}. Causal faithfulness imposes that d-separation is also a necessary condition, such that two variables are statistically independent if and only if d-separation holds. As discussed in \citet{spirtes2000causation}, this implies that multiple causal mechanisms or associations between two variables must not offset each other exactly, as this would lead to statistical independence despite the presence of multiple causal associations.

The reason why we need to impose that conditional statistical independence implies and is implied by the absence of a conditional causal relationship between variables is that, without this assumption, our testing approach might lack power. Specifically, our test verifies conditional statistical independence between the instrument and the outcome, which could hold even if there are multiple causal associations between the variables, namely if those associations exactly offset each other. One example for a failure of faithfulness arises when a nonzero causal effect of one variable on another is exactly canceled out by a confounding association due to unobserved variables affecting both. Conditional on a third (observed) variable, this violation of causal faithfulness implies that the causal and confounding effects cancel out for any value of that variable, thereby imposing strong restrictions on the generality of the causal framework. See, for instance, \citet{Weinberger2017} for a critical discussion of alleged counterexamples to causal faithfulness that are inconsistent with standard causal models. This demonstrates that causal faithfulness, also referred to as stability \citet{Pearl00}, is only violated in very particular parametric models, but not in more general (nonparametric) frameworks. In the context of Figure \ref{causalchain}, a failure of causal faithfulness occurs if, for example, conditional on \( X \), \( Z \) has a causal effect on \( D \), but unobserved confounders of \( Z \) and \( D \) fully offset this effect for any value of \( X \), resulting in statistical independence between \( Z \) and \( D \).

Our second assumption requires that for any values of $X$ in the population, any possible combination of treatment and instrument values exists, which is known as common support.
\begin{assumption}[Common support]\label{asscommonsupport}
\begin{eqnarray*}
\Pr(D=d, Z=z|X)>0 \quad \forall d \in \mathcal{D}\textrm{ and } z \in \mathcal{Z}.
\end{eqnarray*}
\end{assumption}
\noindent Assuming discretely distributed treatments and instruments, Assumption \ref{asscommonsupport} implies that the joint probabilities of any $D=d$ and $Z=z$ conditional on $X$ are larger than zero. In the case of continuously distributed treatments and/or instruments, the joint probabilities are to be replaced by joint density functions conditional on $X$. By applying basic probability theory, $\Pr(D=d, Z=z|X)=\Pr(Z=z|D=d,X)\cdot\Pr(D=d|X)$, where $\Pr(D=d|X)$ is the conditional treatment probability given $X$, known as treatment propensity score, and $\Pr(Z=z|D=d,X)$ is conditional instrument probability given $D,X$, i.e.\ the instrument propensity score. Therefore, Assumption \ref{asscommonsupport} requires that both the treatment and instrument propensity scores are larger than zero, i.e.\ $\Pr(D=d|X)>0$ and $\Pr(Z=z|D,X)>0$ $\forall d \in \mathcal{D}$ and $z \in \mathcal{Z}$.
Common support in the treatment propensity score ensures that causal effects are identified conditional on any value of $X$ in the population, which is a precondition that aggregate treatment effects like the average treatment effect (ATE) given by $E[Y(d)-Y(d')]$ for any $d\neq d'$, are well defined. Under a violation of common support in the treatment propensity score for some values of $X$, one may identify causal effects only for subpopulations whose covariate values satisfy $\Pr(D=d|X)>0$. Common support in the instrument propensity score ensures that the testable implication suggested further below can be verified at any value of $X$ in the population and for any treatment value $d$. A violation of common support in the instrument propensity score implies that one may test the implication only among those covariate and treatment combinations satisfying $\Pr(Z=z|D,X)>0$. Assumption \ref{asscommonsupport} is therefore strictly speaking not required for implementing our testing approach yet to be defined, if we contend ourselves with considering a subpopulation satisfying common support, which may, however, come with the potential caveat of reduced testing power.

Our third assumption requires the treatment and the instrument to be statistically dependent conditional on the covariates, where $\not\!\perp\!\!\!\perp$ denotes statistical dependence.
\begin{assumption}[Conditional dependence between the $D$ and $Z$]\label{ass3}
\begin{eqnarray*}
D \not\!\perp\!\!\!\perp Z|X.
\end{eqnarray*}
\end{assumption}
\noindent Together with Assumption \ref{ass0}, which rules out effects of $D$ on $Z$, Assumption \ref{ass3} either implies that $Z$ causally affects $D$, which is known as first stage effect in the IV literature, or that some (unobserved) characteristics jointly affect $Z$ and $D$ given $X$. This assumption is satisfied in Figure \ref{causalchain}, where the instrument has an impact on the treatment.

Our fourth assumption invokes the conditional independence of the treatment and the potential outcomes given the covariates, with ${\perp\!\!\!\perp}$ denoting statistical independence. This popular assumption in the treatment evaluation literature is also known as selection-on-observables, exogeneity, or unconfoundedness, see e.g.\  \citet{Im04} and \citet{ImWo08}.
\begin{assumption}[Conditional independence of the treatment]\label{ass1}
\begin{eqnarray*}
Y(d) {\perp\!\!\!\perp} D | X\quad \forall d \in \mathcal{D}.
\end{eqnarray*}
\end{assumption}
\noindent Assumption \ref{ass1}, which is referred to as weak unconfoundedness in \citet{Im00} (since it imposes conditional independence for each potential outcome $Y(d)$ separately rather than jointly for all values of $d$ simultaneously), implies that conditional on covariates $X$, there exist no unobserved confounders jointly affecting outcome $Y$ and treatment $D$. This is satisfied in Figure \ref{causalchain}, where unobservables affecting the outcome are independent of the treatment, such that  $U{\perp\!\!\!\perp} D| X$ holds.

Our fifth assumption invokes IV validity, requiring that the instrument is conditionally independent of the potential outcome given the covariates $X$.
\begin{assumption}[Conditional independence of the instrument]\label{ass2}
\begin{eqnarray*}
Y(d) {\perp\!\!\!\perp} Z | X\quad \forall d \in \mathcal{D}.
\end{eqnarray*}
\end{assumption}
\noindent Assumption \ref{ass2} has two implications. First, $Z$ does not directly affect the outcome other than through the treatment conditional on $X$, such that potential outcome is a function of $d$ alone, rather than the instrument, too. For this reason, an IV exclusion restriction holds such that conditional on $X$, $Y(d,z)=Y(d,z')=Y(d)$ for any instrument values $z$ and $z'$, otherwise the conditional independence would be violated. Second, there exist no unobserved confounders jointly affecting $Y$ and $Z$ when controlling for $X$, which is analogous to Assumption \ref{ass1}, but now concerning the instrument rather than the treatment. We note that Assumption \ref{ass2} is not sufficient for identifying causal effects based on the instrument, like the local average treatment effect (LATE) on the subpopulation whose treatment reacts to (or complies with) the instrument, see \cite{Imbens+94} and \cite{Angrist+96}. The IV-based assessment of the LATE hinges on further assumptions, like e.g.\ the (conditional) monotonicity of $D$ in $Z$, which we do not consider here, because it is not required for our identification test.

\begin{figure}[!htp]
	\centering \caption{\label{causalchain2}  Causal graph violating Assumption \ref{ass2}}\bigskip
\centering \includegraphics[width=1\textwidth]{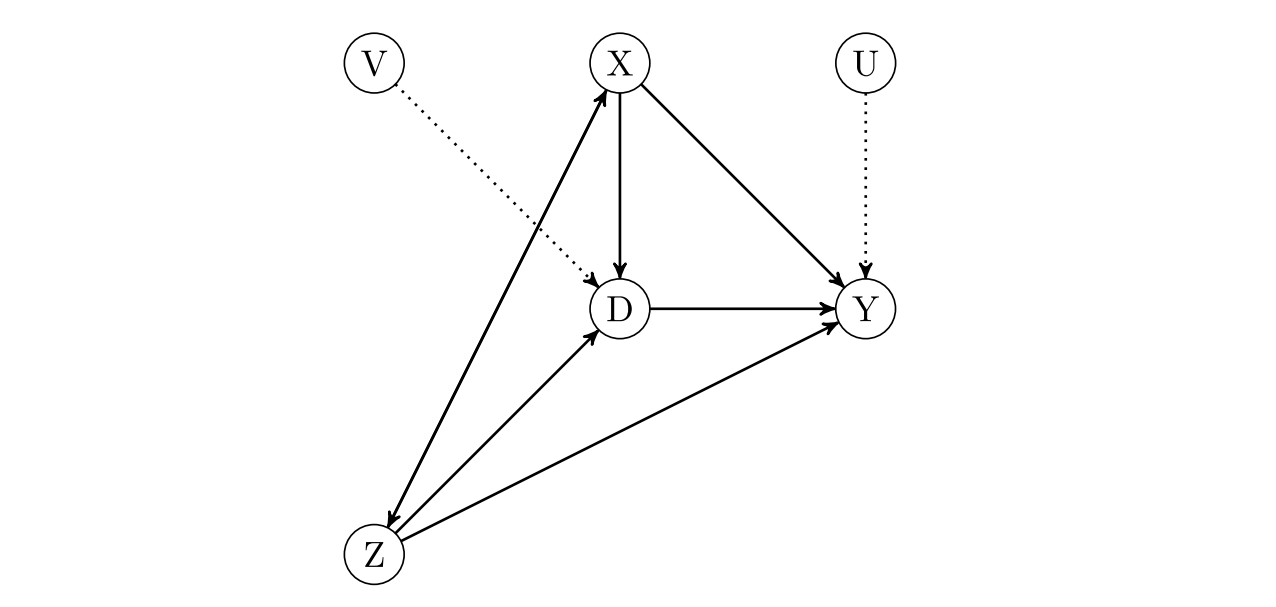}	
\end{figure}

To provide an example where Assumption \ref{ass2} is violated, consider the following nonparametric outcome model: \(Y = \psi(D, X, Z, U)\), where \(\psi\) is an unknown function, and \(U\) denotes unobserved variables affecting the outcome. In this case, \(Z\) directly affects the outcome, as shown in Figure \ref{causalchain2}. Here, the potential outcome \(Y(d) = \psi(d, X, Z, U)\) is not independent of \(Z\) conditional on \(X\). Now, suppose the outcome model is alternatively given by \(Y = \psi(D, X, U)\), so that \(Y(d) = \psi(d, X, U)\). In this case, \(Z\) no longer directly affects \(Y\), and Assumption \ref{ass2} is satisfied if \(U \perp\!\!\!\perp Z | X\), meaning the unobservables and the instrument are conditionally independent given \(X\). This condition holds in Figure \ref{causalchain}, where \(Z\) is a conditionally valid instrument for \(D\), as it neither directly shifts the outcome nor is associated with unobservables affecting the outcome.

In this context, it is worth highlighting the role of causal faithfulness as imposed in Assumption \ref{ass0} once again. If causal faithfulness is violated, the conditional independence between \(Z\) and \(Y\) given \(D\) and \(X\), as tested by our approach (see Theorem \ref{theorem1} below), might hold even if both \(Z\) directly affects \(Y\) and \(U \perp\!\!\!\perp Z | D, X\) fails, because either \(U \perp\!\!\!\perp Z | X\) or \(U \perp\!\!\!\perp D | X\) fails. This can occur if both direct and confounding paths cancel each other out exactly for each value of \(X\). In such a scenario, our test would lack the power to detect such violations of Assumptions \ref{ass1} and \ref{ass2}.

Theorem \ref{theorem1} states the main result of our paper. We show that conditional on Assumptions \ref{ass0} and \ref{ass3}, $Y {\perp\!\!\!\perp}  Z | D=d, X$ is a necessary and sufficient condition for the joint satisfaction of $Y(d) {\perp\!\!\!\perp}  D |  X$ and  $Y(d) {\perp\!\!\!\perp}  Z |  X$. The latter two conditions correspond to Assumptions \ref{ass1} and \ref{ass2} when considering potential outcomes $Y(d)$ that match the factual
treatment assignment $D=d$. Both assumptions are stated as so-called marginal single-world conditions, requiring independence for each potential outcome $Y(d)$ separately rather than jointly (weak unconfoundedness in the sense of \citealt{Im00}), and therefore fall within the single world intervention graphs (SWIG) framework of \citet{richardson2013single} under the Finest Fully Randomized Causally Interpretable Structured Tree Graph (FFRCISTG) semantics of \citet{Ro86}. This framework, and more specifically Assumption~\ref{ass1}, permits identification of the marginal distribution of $Y(d)$ in the population based on observations with $D=d$ when controlling for $X$ and subsequently averaging over the distribution of $X$ in the population. This in turn implies identification of causal effects on potential outcome distributions such as the quantile treatment effect (QTE), see for instance \citet{fir07}. The proof of Theorem~\ref{theorem1} is given in Appendix~\ref{proofth1}.

\begin{theorem}\label{theorem1}
Conditional on Assumptions \ref{ass0} and \ref{ass3}, it holds that
\begin{eqnarray}\label{mainresult}
Y(d) {\perp\!\!\!\perp}  D |  X,\quad Y(d) {\perp\!\!\!\perp}  Z |  X \iff  Y {\perp\!\!\!\perp}  Z | D=d, X \quad\forall d \in \mathcal{D}.
\end{eqnarray}
\end{theorem}


\section{Conditional mean independence}\label{meanind}

The conditional independence assumptions \ref{ass1} and \ref{ass2} refer to the entire distributions of potential outcomes. For assessing the average treatment effect (ATE), we may consider somewhat weaker conditional independence assumptions w.r.t.\ the means of potential outcomes. More formally, we may replace Assumption \ref{ass1} by
the following condition.
\begin{assumption}[Conditional mean independence of the treatment]\label{ass4}
\begin{eqnarray*}
E[Y(d)| D, X]=E[Y(d)|  X] \quad\forall d \in \mathcal{D}.
\end{eqnarray*}
\end{assumption}
\noindent Assumption \ref{ass4} is weaker than Assumption \ref{ass1} as it only imposes the conditional mean independence of $Y(d)$ and $D$ given $X$, but not that of other moments. Analogously, we can weaken Assumption \ref{ass2} to conditional mean independence of the instrument, see Assumption \ref{ass5}.
\begin{assumption}[Conditional mean independence of the instrument]\label{ass5}
\begin{eqnarray*}
E[Y(d)| Z, X]=E[Y(d)|  X] \quad\forall d \in \mathcal{D}.
\end{eqnarray*}
\end{assumption}

When considering Assumptions \ref{ass4} and \ref{ass5} rather than \ref{ass1} and \ref{ass2}, Assumption \ref{ass3} on the conditional dependence of $Z$ and $D$ needs to be replaced, too, by conditional mean dependence.
\begin{assumption}[Conditional mean dependence between the $D$ and $Z$]\label{ass6}
\begin{eqnarray*}
E[D|Z,X] \neq E[D|X].
\end{eqnarray*}
\end{assumption}
\noindent Assumption \ref{ass6} imposes a nonzero first-stage effect of $Z$ in a regression of $D$ on a constant, $Z$, and $X$. It is stronger than the previously imposed Assumption \ref{ass3}, because it requires that the conditional dependence of $D$ and $Z$ necessarily affects the mean of these variables. Any conditional dependence in other moments (like the variance) is irrelevant for this assumption.

Theorem \ref{theorem2} states that conditional on Assumptions \ref{ass0} and \ref{ass6}, the conditional mean independence of $Y$ and $Z$ given $D=d$ and $X$ implies that and is implied by $E[Y(d)| D=d, X]=E[Y(d)|  X]$ and $E[Y(d)| Z, X]=E[Y(d)|  X]$. Therefore, Assumptions \ref{ass4} and \ref{ass5} are testable for verifying the identification of mean potential outcomes and the ATE. The proof is provided in Appendix \ref{proofth2}.
\begin{theorem}\label{theorem2}
Conditional on Assumptions \ref{ass0} and \ref{ass6}, it holds that
\begin{eqnarray}\label{mainresult2}
&&E[Y(d)| D, X]=E[Y(d)|  X],\quad E[Y(d)| Z, X]=E[Y(d)|  X]\\ &\iff&  E[Y|Z,D=d,X]=E[Y|D=d,X] \quad\forall d \in \mathcal{D}.\notag
\end{eqnarray}
\end{theorem}


\section{Testing Identification}\label{DRtest}
First, this section presents the hypotheses for testing the conditional mean independence in \eqref{mainresult2}, which implies the identification of average effects. We assume that the supposed instrument $Z$ to be a scalar, while $X$ is a vector of observed covariates.
To ease notation, we henceforth denote the conditional mean outcome as $\mu(z,d,x)=E[Y|Z=z,D=d,X=x]$. Throughout this section, we use $W=(Y,Z,D,X)$ to denote the vector of all observed variables for a given observation. Considering a discretely distributed instrument $Z$, the conditional mean independence \eqref{mainresult2} is equivalent to the following null hypothesis $H_0$:
\begin{eqnarray}\label{nullhyp1}
\mu(z,d,x)-\mu(z',d,x)=0 \quad\forall z\neq z' \in \mathcal{Z}, d \in  \mathcal{D},\textrm{ and }x \in \mathcal{X}.
\end{eqnarray}
Under the null hypothesis, the mean conditional outcome is constant across $Z$ given any value of $D$ and $X$. For a binary instrument, which we assume in the following, equation \eqref{nullhyp1}, for instance, corresponds to $$\mu(1,d,x)-\mu(0,d,x)=0.$$
Under a linear regression model for the outcome with homogeneous effects, we can test the null hypothesis \eqref{nullhyp1} 
based on regressing $Y$ on a constant, $Z$, $D$, and $X$ and verifying whether the coefficient on $Z$ is statistically significantly different from zero. However, when assuming a more flexible nonparametric model, the conditional statistical association of $Z$ and $Y$ is allowed to be heterogeneous across different values of $Z$, $D$, and $X$. In theory, one then might want to verify the respective null hypothesis at all values of the instrument, the treatment, and the covariates, to check for any possible violation. However, if some or all of these variables are continuously distributed, this implies an infinite number of hypotheses to be tested. Even under (mostly) discretely distributed variables, statistical power in finite samples quickly decreases in the conditioning set as a function of the dimension and support of $X$. 
Therefore, a test statistic must necessarily involve an aggregate measure over its domain (or support), see e.g. \cite{racine1997consistent} and \cite{racine2006testing} who provide a nonparametric significance test based on an aggregate $L_2$-norm that is related to our approach in Section \ref{squareddiff}.
Another way to circumvent such issues of limited finite sample power and multiple hypothesis testing is to verify whether the null hypothesis holds on average. For a binary instrument as considered in our application, this amounts to testing the following condition:
\begin{eqnarray}\label{nullhyp3}
H_0: E[\mu(1,D,X)-\mu(0,D,X)]=0.
\end{eqnarray}
We henceforth denote this average difference in conditional means by $\Delta=E[\mu(1,D,X)-\mu(0,D,X)]$. $\Delta$ may be estimated by treatment evaluation methods for assessing the ATE, such as propensity score matching, see \citet{rosenbaum1983} and \citet{RosenbaumRubin1985}, inverse probability weighting (IPW), see \citet{Horvitz52} and \citet{Hirano+00}, or doubly robust (DR) methods, see \citet{Robins+94} and \citet{RoRo95}. 


\subsection{Testing based on doubly robust estimation}\label{DRtestH}

We henceforth suggest a testing approach based on doubly robust (DR) estimation.
To this end, denote $p(D,X):=E[Z|D,X]=\Pr(Z=1|D,X)$ the conditional instrument probability or instrument propensity score.
The DR approach exploits both propensity scores and conditional means to estimate the following expression for $\Delta$ (which is equivalent to $E[\mu(1,D,X)-\mu(0,D,X)]$): 
\begin{eqnarray}\label{drselobs}
\Delta&=&E\left[\phi(D,X) \right]
\end{eqnarray}
with
\begin{align}\label{DR}
\phi(D,X)&=\mu(1,D,X)-\mu(0,D,X)+\frac{(Y-\mu(1,D,X)) Z}{p(D,X)}
-\frac{(Y-\mu(0,D,X)) (1-Z)}{1-p(D,X)}.
\end{align}
Here, $\phi(D,X)$ is the efficient (and Neyman-orthogonal) score function, into which $\mu(Z,D,X)$ and $p(D,X)$ enter as first-step or nuisance parameters. The sample analog of \eqref{drselobs} consistently estimates $\Delta$ if either the model for $\mu(Z,D,X)$ or for $p(D,X)$ is correctly specified, which is known as DR property.
In particular when the set of covariates $X$ is large, one might estimate $\mu(Z,D,X)$ and $p(D,X)$ by machine learning (ML) algorithms, in order to control for important confounders in a data-driven way (rather than using an ad-hoc rule which may introduce pre-testing issues). This double machine learning (DML) approach is typically combined with cross-fitting, which consists of estimating the the nuisance parameter models and the score function in non-overlapping subsets of the data, with the roles of the subsets for the estimation steps being sequentially swapped. As no observation enters both estimation steps at the same time, cross-fitting avoids correlations between the estimation of the models of $\mu(Z,D,X)$ and  $p(D,X)$ on the one hand and of the score function $\phi(D,X)$  on the other hand and thus, overfitting bias. Finally, taking the sample average of the estimated score function yields an estimate of $\Delta$, in analogy to the population average in \eqref{drselobs}. Under specific regularity conditions, e.g.\ the convergence rate of ML-based estimators of $\mu(Z,D,X)$ and $p(D,X)$ being faster than $n^{1/4}$, DML is root-$n$-consistent, see the discussion in \cite{doubleML}.

However, verifying the condition \eqref{nullhyp3} rather than \eqref{nullhyp1} has its caveats. By averaging over values of $D$ and $X$ (and $Z$, if it is non-binary) when testing the null hypothesis, there is a risk of averaging out violations of the conditional mean independence \eqref{mainresult2} such that they cannot be detected. By aiming at increasing finite sample power through averaging, we sacrifice asymptotic power as we do not test the hypotheses separately for distinct values of our conditioning set. To more optimally trade off asymptotic and finite sample power, we outline a further testing approach, which borrows from the literature on estimating conditional average treatment effects (CATE) and investigating effect heterogeneity across observed characteristics based on ML, see e.g.\  \citet{WagerAthey2018}. Denoting the conditional mean difference $\mu(1,d,x)-\mu(0,d,x)$ by $\Delta(d,x)$, we use the score function $\phi(D,X)$ for assessing whether $\Delta(d,x)$ is heterogeneous across values of $(D,X)$ and thus necessarily non-zero for some values of the conditioning set. 
A practical issue is that we would like to focus on those variables in the set $(D,X)$ which importantly predict the effect heterogeneity of $\Delta(D,X)$ and thus, drive the statistical power of our test. This suggests the use of ML for determining the crucial predictors of the estimate of $\phi(D,X)$. 

However, if the ML-based detection of important predictors of $\Delta(d,x)$ and hypothesis testing e.g.\ of \eqref{nullhyp1} based on those predictors proceeds in the very same data, this may entail overfitting bias. This can entail spurious rejections of the null hypothesis due to a correlation of the estimation steps of predictor selection and testing, in analogy to the discussion in \citet{AtheyImbens2016}. For this reason, we apply a sample splitting approach which avoids such correlations, see e.g.\ \citet{leeetal2020}, by randomly partitioning our sample into two non-overlapping subsamples. In the first subsample, we apply the previously discussed DML procedure to estimate the efficient score function, henceforth denoted by $\hat{\phi}(D,X)$. Still in the first subsample, $\hat{\phi}(D,X)$ is predicted as a function of the predictors $(D,X)$ using ML, for instance based on a so-called decision tree, see \citet{MorganSonquist1963} and \citet{Breimanetal1984}.  A decision tree consists of recursively splitting the covariate space into subsets (or leaves), such that predictive power w.r.t.\ $\hat{\phi}(D,X)$ is maximized. In the second subsample, we first estimate $\hat{\phi}(D,X)$ based on DML and then conduct the hypotheses tests conditional on the covariates that have been found to importantly predict the score in the first subsample.

More formally, let $n$ denote the total sample size (i.e.\ the sum of observations in the first and second subsample) and $i$ be the index of observations in the sample, i.e.\ $i\in \{1,...,n\}$. Furthermore, denote by $L_m$ a specific leaf or subset and by $M$ the total number of leaves defined by the regression tree (or any other machine learning algorithm) in the first subsample, such that $m\in \{1,...,M\}$.
The average of $\Delta(D,X)$ within the respective leaf $L_m$ is denoted by $\Delta(L_m)=E[\Delta(D,X)|D,X \in L_m].$
Hypothesis testing amounts to verifying whether the averages in the various leaves are statistically significantly different from zero. A leaf-specific null hypotheses is defined as follows:
\begin{eqnarray}\label{nullhyp3a}
H_0: E[\mu(1,D,X)-\mu(0,D,X)| (D,X) \in L_m]=0.
\end{eqnarray}

Under specific regularity conditions, the estimated parameters are root-n-consistent and asymptotically normally distributed with a variance which is not affected by the fact that $\hat{\phi}(D_i,X_i)$ is a ML-based estimate of the true (but unknown) score function $\phi(D_i,X_i)$. As discussed in \citet{SemenovaChernozhukov2020}, such a favorable behavior can be attained if the ML-based estimators of the nuisance parameters converge to the respective true models with a rate faster than $n^{1/4}$ and if the number of leaves $M$ is small relative to the sample size $n$. In this case, we can directly consider the t-statistics for testing the null hypothesis in specific leaves. To account for multiple hypothesis testing issues due to running the test in several leaves, we may apply a statistical correction which controls for the expected proportion of spurious rejections among the rejected hypotheses, the so-called false discovery rate, as e.g.\ suggested \citet{BenjaminiHochberg1995}, or the standard Bonferroni correction. Alternatively, we can run an F-test for the joint satisfaction of the null hypothesis \eqref{nullhyp3a} across all leaves $L_m$ if we assume a linear form of the regression function $\mu$.

One drawback of such corrections for multiple testing is that they can drastically reduce testing power if the number of leaves $M$ is non-negligible. To mitigate this issue, we base testing on uniformly valid confidence intervals for all target parameters $\theta_m=E[(\mu(1,D,X)-\mu(0,D,X))I\{ D,X \in L_m \}]$, $m=1,\dots,M$ obtained by the multiplier bootstrap. The target parameter $\theta_m$ equals zero if and only if $E[\mu(1,D,X)-\mu(0,D,X)| (D,X) \in L_m]=0$, given that $P(X,D\in L_m)>0$. To construct uniformly valid confidence intervals, we consider the $M$-dimensional score function $\phi^{M}=(\phi_1,\dots,\phi_M)$ with
\begin{align}\label{DR_sim}
\phi_m(W_i,\theta,\eta)=&\Bigg(\mu(1,D_i,X_i)-\mu(0,D_i,X_i)
+\frac{(Y_i-\mu(1,D_i,X_i))\cdot Z_i}{p(D_i,X_i)}\nonumber\\
&-\frac{(Y_i-\mu(0,D_i,X_i))\cdot (1-Z_i)}{1-p(D_i,X_i)}\Bigg)
I\{ D_i,X_i \in L_m \}-\theta_m
\end{align}
for all $m=1,\dots,M$ with $\eta=(\eta_1,\eta_2)=(\mu,p)$.

To derive the large sample behavior of this approach, we introduce further notation. Let $\delta_N$ be sequences of positive constants approaching $0$. Furthermore, let $C$ and $q$ be fixed strictly positive constants with $q>2$. In the following, we rely on cross-fitted DML and split the $N$ observations in the second subsample (used for testing) into $K$ folds $I_k$, $k=1,\dots,K$, of size $N/K$ as described in Definition 3.2 in \cite{doubleML}. For simplicity, assume
that $N/K$ is an integer. For each $k\in [K]$, we obtain
\begin{align*}
\hat{\eta}_k=\left(\hat{\mu}((W_i)_{i\in I_{k}^c}),\hat{p}((W_i)_{i\in I_{k}^c}))\right),
\end{align*}
which is an ML-based estimate of $\eta_0$. We then estimate the target parameter $\theta_M$ by $\hat{\theta}_M$, which solves
\begin{align*}
\frac{1}{K}\sum_{k=1}^K\mathbb{E}_{N,k}[\phi^M(W,\hat{\theta},\hat{\eta}_k)]=0,
\end{align*}
where $\mathbb{E}_{N,k}$ is the empirical expectation over the kth fold of the data. We now impose a set of regularity conditions required for the construction of uniform confidence intervals and the asymptotic normality of our testing approach, which is stated in Theorem \ref{theoremDR}. 
\begin{assumption}[Uniform confidence intervals]\label{assDR}
Define $U:=Y-\mu(Z,D,X)$, and let $c$ and $C$ denote strictly positive constants that are independent of $n$ and $P\in\mathcal{P}$. The following assumptions need to hold for all $n\ge 3$, $P\in\mathcal{P}$ and $q>2$, where $\mathcal{P}$ denotes the set of all probability distributions compatible with the assumptions of the model:
\begin{itemize}
\item[i)] It holds $\|Y\|_{P,q}<C$, $\mathbb{E}\left[U^2 I\{D,X\in L_m\}\right]>c$ for all $m=1,\dots,M$\\
and  $\|\mathbb{E}[U^2|D,X]\|_{P,\infty}<C$.
\item[ii)] It holds, $\epsilon<P(Z=1|X,D\in L_m)<1-\epsilon$ for all $m=1,\dots,M$.
\item[iii)] Given a random subset $I$ of $[N]$ of size $n = N/K$, the nuisance parameter estimator $\hat{\eta}_0=\hat{\eta}_0((W_i)_{i\in I^c})$ obeys the following conditions: With $P$-probability not less than $1-o(1)$, $\max(\|\hat{\eta}_1-\eta_{0,1}\|_{P,q},\|\hat{\eta}_2-\eta_{0,2}\|_{P,q})\le C$,  $\max(\|\hat{\eta}_1-\eta_{0,1}\|_{P,2},\|\hat{\eta}_2-\eta_{0,2}\|_{P,2})\le \delta_N$,
$\|\hat{\eta}_2-1/2\|_{P,\infty}\le 1/2-\epsilon$ and $\|\hat{\eta}_1-\eta_{0,1}\|_{P,2} \times \|\hat{\eta}_2-\eta_{0,2}\|_{P,2}\le\delta_NN^{-1/2}$.
\end{itemize}
\end{assumption}
\begin{theorem}\label{theoremDR}
Conditional on Assumptions \ref{assDR}, under $H_0$, it holds that
\begin{eqnarray}
\sqrt{N}\Sigma^{-1}\left(\hat{\theta}_M-\theta_M\right)\rightsquigarrow N(0,I_M)
\end{eqnarray}
uniformly over $P\in\mathcal{P}$, where $\Sigma^2 = \mathbb{E}[\phi^M(W,\theta_0,\eta_0)\phi^M(W,\theta_0,\eta_0)']$ with diagonal elements $\sigma_m^2$, $m=1,\dots,M$. Moreover, the result continues to hold if
$\Sigma^2$ is replaced by
\begin{align*}
\hat{\Sigma}^2:=\frac{1}{K}\sum_{k=1}^K\mathbb{E}_{n,k}\left[\phi^M(W,\hat{\theta}_M,\hat{\eta}_k)\phi^M(W,\hat{\theta}_M,\hat{\eta}_k)'\right].
\end{align*}
\end{theorem}
The proof of Theorem \ref{theoremDR} is provided in Appendix \ref{proofDR} and we note that its result also holds under the alternative hypothesis $H_1$ if the variance $\Sigma^2$ of the score is non-degenerate. 
Theorem \ref{theoremDR} can be used to construct confidence regions for any scalar parameter $l'\theta_M$ for some $ M \times 1$ vector $l$. Using for instance the multiplier bootstrap, we can also construct a confidence interval for $\sup_{m=1,\dots,M}|\theta_m|$ to simultaneously test the hypotheses \eqref{nullhyp3a} for all $m=1,\dots,M$. To this end, we define the process
\begin{align*}
\hat{\mathcal{G}}=\left(\hat{\mathcal{G}}_m\right)_{m=1,\dots,M}=\left(\frac{1}{\sqrt{N}}\sum_{i=1}^N\zeta_i\hat{\sigma}^{-1}_m\phi_m(W_i,\hat{\theta}_m,\hat{\eta})\right)_{m=1,\dots,M},
\end{align*}
where $(\zeta_i)_{i=1}^N$ are standard normal random variables, which are independent of each other and of the data $(W_i)_{i=1}^N$.  $\hat{\sigma}_m^{-1}$, $m=1,\dots,M$, are the diagonal elements of $\hat{\Sigma}^{-1}$. The critical value $c_\alpha$ obtained through the multiplier bootstrap corresponds to the ($1-\alpha$)-quantile of the conditional distribution of $\sup_{m=1,\dots,M} |\hat{\mathcal{G}}_m|$ given the data $(W_i)_{i=1}^N$. The null hypothesis is rejected if
$
\sup_{m=1,\dots,M}\sqrt{N}\hat{\sigma}^{-1}_m|\theta_m| > c_{\alpha}.
$
It is worth noting that one can also construct hypothesis tests based on the $l_p$-norm, $p\ge 1$, rather than the infinity norm of $\{\hat{\sigma}^{-1}_m\theta_m\}_{m=1}^M$. In this case, we reject \eqref{nullhyp3a} if
\begin{align}\label{p-norm}
\sqrt{N}\|\{\hat{\sigma}^{-1}_m\theta_m\}_{m=1}^M\|_p> c_{\alpha}^{(p)},
\end{align}
with $c_{\alpha}^{(p)}$ being the ($1-\alpha$)-quantile of the conditional distribution of $\|\{\hat{\mathcal{G}}_m\}_{m=1}^M\|_p$ given the data $(W_i)_{i=1}^N$. This can lead to more efficient confidence intervals with lower volume compared to standard multiple testing approaches and hence to tests with higher power, see e.g.\  \cite{https://doi.org/10.48550/arxiv.2105.09028} and \cite{10.1093/biomet/asac030}. 

Finally, we note that since Theorem \ref{theoremDR} also holds for any $\Delta\neq 0$, we can reverse the role of the null and alternative hypotheses to construct hypotheses tests of the form
\begin{eqnarray}\label{nullhyp4}
H_0: \sqrt{N}\|\{\hat{\sigma}^{-1}_m\theta_m\}_{m=1}^M\|_p > \Delta_0,
\end{eqnarray}
for a predefined threshold $\Delta_0>0$. In such a framework as e.g.\ advocated by \cite{bilinski2018nothing}, the null hypothesis $H_0$ postulates a non-negligible violation of the testable implication, while the alternative hypothesis $H_1$ states that the violation is close to zero, i.e.\ smaller than the absolute value of $\Delta_0$. For this reason, a statistically significant test statistic suggests that violations exceeding the threshold $\Delta_0$ can be ruled out in order to justify ATE estimation based on the selection-on-observables assumption.


\subsection{Testing based on squared differences}\label{squareddiff}
In this section, we suggest a further testing approach based on squared differences in conditional mean outcomes. More concisely, we aim at testing for violations of hypothesis \eqref{nullhyp1} globally, i.e.\ across all values of $D$ and $X$, by verifying the null hypothesis $H_0$:
\begin{align}\label{H0_quadratic}
    \theta_0:=\mathbb{E}\left[\left(\mu(1,D,X)-\mu(0,D,X)\right)^2\right]=0.
\end{align}
In contrast to equation \eqref{drselobs}, equation \eqref{H0_quadratic} uses an aggregate $L_2$-type measure to test violations across values of $D$ and $X$, which is a common approach in specification tests for nonparametric regression, see e.g. \cite{racine1997consistent}, \cite{racine2006testing}, \cite{hong1995consistent} and \cite{wooldridge1992test}. Again, we consider ML for estimating $\mu$ to allow for a high-dimensional $X$. For this reason, we test \eqref{H0_quadratic} based on a moment condition which uses the following score function:
\begin{align}\label{score2}
\phi_2(W,\theta,\eta)=(\eta(1,D,X)-\eta(0,D,X))^2-\theta+\zeta,
\end{align}
where $\eta_0(Z,D,X)=\mu(Z,D,X)$ is the true nuisance parameter, and $\zeta$ is an independent mean-zero random variable, which satisfies $\|\zeta\|_{P,q}<C$ for $q>2$ and has a variance  $\sigma_\zeta^2>0$. In contrast to the DR approach in Section \ref{DRtest}, testing exclusively relies on the conditional mean outcome as nuisance parameter, while propensity score estimation is not required. 

The additional random variable $\zeta$ is introduced to avoid a degenerate distribution of our estimator under $H_0$ - a common problem in specification tests, see e.g. \cite{hong1995consistent} and \cite{wooldridge1992test}. The variance $\sigma_\zeta^2$ acts like a tuning parameter and should be chosen as a function of the sample size $n$ . A high variance should ensure that our test based on $\phi_2$ comes close to the nominal level $\alpha$ in small samples. On the negative side, a high variance could reduce testing power by whitewashing violations of the null hypothesis.
We note that another approach avoiding a degenerate distribution under $H_0$ consists of combining the quadratic score with the DR score in \eqref{drselobs}.
To estimate the parameter $\theta_0$, we use cross-fitting and split the data into $K$ subsamples of size $N=n/K$. The cross-fitted estimator is given by
$\hat{\theta}=\frac{1}{K}\sum_{k=1}^K\mathbb{E}_{N,k}[\left(\hat{\eta}_k(0,D_i,X_i)-\hat{\eta}_k(1,D_i,X_i)\right)^2+\zeta_i]$.
The following assumption imposes regularity conditions required for the asymptotic normality of our test under the null hypothesis, as stated in Theorem \ref{theorem3}. 
\begin{assumption}[Asymptotic Normality]\label{ass7}
The following assumption needs to hold for all $n\ge 3$, $P\in\mathcal{P}_n$ and $q>2$, where $\mathcal{P}_n$ denotes a sequence of probability distributions that may change with the sample size $n$ and are compatible with the assumptions of the model: Given a random subset $I$ of $\{1,\dots,n\}$ of size $N = n/K$, the nuisance parameter estimator $\hat{\eta}_0=\hat{\eta}_0((W_i)_{i\in I^c})$ obeys $\|\hat{\eta}-\eta_{0}\|_{P,2q}\le C$,  $\|\hat{\eta}-\eta_{0}\|_{P,4}\le \delta_n$, and $\|\hat{\eta}-\eta_{0}\|_{P,2}\le\delta_n^{1/2}n^{-1/4}$ with $P$-probability not less than $1-o(1)$.
\end{assumption}
\begin{theorem}\label{theorem3}
Conditional on Assumptions \ref{ass7}, under $H_0$, it holds
$$\sqrt{n}\sigma^{-1}\hat{\theta}\rightsquigarrow N(0,1)$$
uniformly over $P\in\mathcal{P}_n$, where 
$\sigma^2 = \mathbb{E}[(\eta_0(1,D,X)-\eta_0(0,D,X))^4]+\sigma_\zeta^2=\sigma_\zeta^2$.
Consequently, a test that rejects the null hypothesis $H_0$ if $\sqrt{n}\hat{\sigma}^{-1}|\hat{\theta}|>\Phi^{-1}(1-\alpha/2)$ has asymptotic level $\alpha$ where $$\hat{\sigma}^2:=\frac{1}{K}\sum_{k=1}^K\mathbb{E}_{N,k}\left[\left(\hat{\eta}_k(1,D_i,X_i)-\hat{\eta}_k(0,D_i,X_i)\right)^4\right]+\sigma_\zeta^2.$$
\end{theorem}
It is worth noting that Theorem \ref{theorem3} only holds under $H_0$, because Neyman orthogonality is only satisfied under $H_0$, as shown in the proof of Theorem \ref{theorem3} in Appendix \ref{proofth3}. To provide theoretical power results under $H_1$, one could orthogonalize the score \eqref{score2} using the methodology of \cite{chernozhukov2015post} and \cite{doubleML}. This is an open question for future research.


\section{Simulation study}\label{simulation}
This section provides a simulation study to investigate the finite sample behavior of our testing approaches introduced in Section \ref{DRtest} based on the following data generating process:
\begin{eqnarray*}
	Y &=& D + X'\beta + \gamma Z + \delta W  + U,\\
	D &=& I\{X'\beta + Z + W  + V >0\},\\
	X &\sim& N(0,\sigma^2_X), Z \sim bernoulli(0.5),\\
W&\sim& N(0,\sigma^2_W),  U\sim N(0,\sigma^2_U), V\sim N(0,\sigma^2_V),
\end{eqnarray*}
 with $X$, $Z$, $W$, $U$, $V$ being independent of each other.
Outcome $Y$ is a linear function of $D$ (whose treatment effect is one), covariates $X$ (for $\beta\neq 0$), the unobservables $W$ (for $\delta\neq 0$) and $U$, and the supposed instrument $Z$ if the coefficient $\gamma\neq 0$. The binary treatment $D$ is a function of $X$, $Z$, and the unobservables $W$ and $V$. While the supposed instrument $Z$ is binary, the unobserved terms $U,V,W$ are normally distributed random variables that are independent of each other, of $Z$, and of $X$ with $\sigma_U=\sigma_V=0.1$ and $\sigma_W=0.25$. $X$ is a vector of covariates which follow a normal distribution with a zero mean and a covariance matrix $\sigma^2_X$ that is obtained by setting the covariance of the $i$th and $j$th covariate in $X$ to $0.5^{|i-j|}$. The coefficient vector $\beta$ gauges the effects of the covariates on $Y$ and $D$, respectively, and thus, the magnitude of confounding due to observables. The $i$th element of the coefficient vector $\beta$ is set to $0.7/i$ for $i=1,...,p$, implying a linear decay of covariate importance in terms of confounding.

We analyze the performance of our testing approach in $1000$ simulations under two sample sizes of $n=1000$ and $4000$ when setting the number of covariates to $50$. We estimate $\Delta$, respectively $\theta_0$, based on DML with 3-fold cross-fitting using lasso regression, see \citet{Tibshirani96}, as ML method for estimating the nuisance parameters, i.e.\ linear and logit specifications of the outcome and treatment equations with cross-validated lasso penalty terms. Observations whose instrument propensity scores are close to zero, namely smaller than a threshold of $0.01$ (or 1\%), are dropped from the estimation in order to avoid an explosion of the propensity score-based weights, which might heavily increase the variance of estimating $\Delta$. 
We also analyze the test performance when using the score based on the squared difference in equation (15)
to estimate $\theta_0$ in (14).
In this case, we choose $\zeta\sim \mathcal{N}(0,\sigma_\zeta^2)$ where the variance term $\sigma_\zeta$ decreases in the sample size, by setting $\sigma_\zeta=500/n$ in (15). We note that the tests based on average and squared violations in the total sample are available in the `identificationDML' function of the `causalweight' package by  \citet{BodoryHuber2018} for the statistical software \textsf{R}, which makes use of the `SuperLearner' package by \cite{vanderLaanetal2007} to select ML algorithms for nuisance parameter estimation.

\begin{table}[htbp]
\begin{center}
	\caption{Simulations}
	\label{tab:sim}
	\begin{tabular}{c|cccc|cccc}
  \hline\hline
	&\multicolumn{4}{c}{Test based on $\Delta$} & \multicolumn{4}{c}{Test based on $\theta$}\\
	 \hline
 sample size & est & std & mean se & rej. rate & est & std & mean se & rej. rate \\
	\hline
   &\multicolumn{8}{c}{Assumptions \ref{ass4} and \ref{ass5} hold ($\delta=0$, $\gamma=0$)} \\
   \hline
1000 & -0.0030  & 0.0069  & 0.0066  & 0.151 & 0.0034  & 0.0152  & 0.0158  & 0.097 \\
4000 & 0.0016  & 0.0034 & 0.0033  & 0.135 & 0.0009 & 0.0077 & 0.0079  & 0.091\\
   \hline
   &\multicolumn{8}{c}{Ass.\ \ref{ass4} violated, Ass.\ \ref{ass5} holds ($\delta=2$, $\gamma=0$)} \\
   \hline
  1000 & -0.0695  & 0.0367 & 0.0328  & 0.657 & 0.0721 &  0.0192 &  0.0162 & 0.992\\
  4000 & -0.0613 & 0.0170  & 0.0166  & 0.979 & 0.0232 & 0.0040 &  0.0020 & 1.000 \\
 \hline
  &\multicolumn{8}{c}{Ass.\ \ref{ass4} holds, Ass.\ \ref{ass5} violated ($\delta=0$, $\gamma=0.1$)} \\
   \hline
   1000  & 0.0970 & 0.0069  & 0.0066 & 1.000 & 0.0126  & 0.0152  & 0.0158 & 0.186 \\
   4000 & 0.0984 & 0.0034 & 0.0033 & 1.000 & 0.0101 & 0.0020  & 0.0020 & 1.000\\
   \hline
\end{tabular}
\end{center}
\par
{\scriptsize Notes: columns `est', `std', and  `mean se' provide the average estimate of $\Delta$ and $\theta$, respectively, its standard deviation, and the average of the estimated standard error across all samples. `rej. rate' gives the empirical rejection rate when setting the level of statistical significance to 0.1 (or 10\%). }
\end{table}

Table \ref{tab:sim} reports the simulation results. The top panel focusses on the case that $\delta=\gamma=0$, such that both Assumptions \ref{ass4} and \ref{ass5} are satisfied. Already under the smaller sample size of $n=1000$, the average estimate of $\Delta$ (`est') across all simulations is close to zero, and quickly approaches this true value as the sample size increases. Accordingly, the empirical rejection rate amounts to 0.151 or 15.1\% under the smaller sample size, which is only somewhat higher than the nominal rate of 10\% when setting the level of statistical significance to $0.1$. Under the larger sample size of $n=4000$, the rejection rate corresponds to 13.5\% and, thus, appears to approach the nominal level. Furthermore, the average standard error across all simulations (`mean se') is generally close to the actual standard deviation (`std') of DML. We also see a root-n consistent behavior in the sense that the standard deviation of the estimator is cut by half when the sample size is quadrupled, while the bias is close to zero.

The intermediate panel presents the results when Assumption \ref{ass4} does not hold ($\delta=2$, $\gamma=0$), such that the treatment is not conditionally mean independent. The DML estimates are substantially different from zero, amounting to $-0.0695$ and $-0.0613$ for $n=1000$ and $n=4000$, respectively. Furthermore, the test's statistical power to detect the violation quickly increases in the sample size, with a rejection rate amounting to 65.7\% under the lower and 97.9\% under the higher sample size. Similar conclusions apply to the violation of Assumption \ref{ass5} ($\delta=0$, $\gamma=0.1$), such that the instrument is not conditionally mean independent. The lower panel of Table \ref{tab:sim} shows that the rejection rates correspond to 100\% for both sample sizes, $n=1000$ and $n=4000$. Summing up, we find the empirical size and power of our test based on $\Delta$ to be very decent in these simulation scenarios.

The empirical size of our test based on the squared difference $\theta$ is generally close to the nominal level of 10\% under the null hypothesis, as indicated in the top panel of Table \ref{tab:sim}. Also the power is generally quite decent. Under a violation of Assumption \ref{ass4} ($\delta=2$, $\gamma=0$), the rejection rate is close to 100\% under either sample size, see the intermediate panel. Under a violation of Assumption \ref{ass5} ($\delta=0$, $\gamma=0.1$), the power of the test based on $\theta$ is quite low for $n=1000$ but increases fast in the sample size, with a rejection rate of 100\% for $n=4000$, see the lower panel. In contrast to the estimation of $\Delta$, we do not observe a root-$n$ consistent behavior of the test based on $\theta$ in the intermediate and lower panel of Table \ref{tab:sim}, which is not surprising as orthogonality only holds under the null hypothesis, see Theorem \ref{theorem3}.

In a next step, we consider a simulation setting with effect heterogeneity:
\begin{eqnarray*}
	Y &=& D + X'\beta + \gamma ZX_1 + \gamma ZX_2 + \delta WX_1+\delta WX_2  + U,\\
	D &=& I\{X'\beta + Z + W  + V >0\},\\
	X &\sim& N(0,\sigma^2_X), Z \sim bernoulli(0.5),\\
W&\sim& N(0,\sigma^2_W),  U\sim N(0,\sigma^2_U), V\sim N(0,\sigma^2_V),
\end{eqnarray*}
 with $X$, $Z$, $W$, $U$, $V$ being independent of each other and $X_1$ and $X_2$ denoting the first and second covariates in $X$, respectively. For $\delta\neq0$ and $\gamma\neq0$, Assumptions \ref{ass4} and \ref{ass5}, respectively, are violated, but in contrast to our previous simulation design, the violations are now heterogeneous in $X_1$ and $X_2$. We also note that the violations cancel out when averaging over $X_1$ and $X_2$, because the covariates are normally distributed and centered around zero.  

\begin{table}[htbp]
\begin{center}
	\caption{Simulations: Effect Heterogeneity}
	\label{tab:sim:het}
	\begin{tabular}{c|cccc|cccc}
  \hline\hline
	&\multicolumn{4}{c}{Test based on $\Delta$} & \multicolumn{4}{c}{Test based on $\theta$}\\
	 \hline
 sample size & est & std & mean se & rej. rate & est & std & mean se & rej. rate \\
  \hline
    &\multicolumn{8}{c}{Assumptions \ref{ass4} violated, \ref{ass5} holds ($\delta=2$, $\gamma=0$)} \\
   \hline
1000 & 0.0015 & 0.0656 & 0.0584  & 0.139 & 0.1738 & 0.0369  & 0.0178  & 1.000 \\
4000 & 0.0181 & 0.0320 &  0.0300 & 0.185 & 0.0582 & 0.0122 & 0.0080 & 1.000 \\
   \hline
   &\multicolumn{8}{c}{Ass.\ \ref{ass4} holds, Ass.\ \ref{ass5} violated ($\delta=0$, $\gamma=0.1$)} \\
   \hline
  1000 & -0.0033 & 0.0089 & 0.0086  & 0.143 & 0.0336 & 0.0155 &  0.0159 & 0.678\\
  4000 & -0.0017 & 0.0044 & 0.0043 & 0.120 & 0.0309 & 0.0077 &  0.0079 & 0.995 \\
 \hline
  &\multicolumn{8}{c}{Ass.\ \ref{ass4} and Ass.\ \ref{ass5} violated ($\delta=2$, $\gamma=0.1$)} \\
   \hline
   1000  & 0.0013 & 0.0657 & 0.0586  & 0.139 & 0.2035 & 0.0519  & 0.0187 & 1.000 \\
   4000  & 0.0181 & 0.0321 & 0.0300 & 0.188 & 0.0864 &  0.0212 & 0.0082 & 1.000 \\
   \hline
\end{tabular}
\end{center}
\par
{\scriptsize Notes: columns `est', `std', and  `mean se' provide the average estimate of $\Delta$ and $\theta$, respectively, its standard deviation, and the average of the estimated standard error across all samples. `rej. rate' gives the empirical rejection rate when setting the level of statistical significance to 0.1 (or 10\%). }
\end{table}

Table \ref{tab:sim:het} provides the results for testing based on $\Delta$ and $\theta$ under a violation of Assumption \ref{ass4} (upper panel), Assumption \ref{ass5} (intermediate panel), and both assumptions (lower panel). The DR estimator of  $\Delta$ has a low power in all settings and under either sample size, due to averaging out violations across values of $X$. In contrast, the estimator based on the squared difference $\theta$ has a very high power in all settings investigated.

Finally, we investigate empirical size and testing power under effect heterogeneity when estimating violations based on DR in subsets of the data defined as a function of variables in $(D,X)$ that importantly predict such violations, as discussed in Section \ref{DRtest}. To this end, we randomly split our data into two halves and use the first subsample to estimate the DR score functions based on cross-fitting and the random forest as implemented in the `grf' package by \citet{TibshiraniAtheyWager2020grf} for nuisance parameter estimation. We then apply yet another random forest for determining the importance of the variables in $(D,X)$ for predicting the estimated scores based on the mean squared error-criterion, using the `randomForest' package by \citet{LiawWiener2002}. According to variable importance, we pick those three variables which explain most of the heterogeneity in violations. In the second subsample, we split the data at the median of these variables, such that we obtain $M=6$ subsets in which we estimate the violation $\theta_m$ based on the DR score function in equation (10). Finally, we control for multiple testing using the Bonferroni correction or the multiplier bootstrap based on the $\|\cdot\|_\infty$ and $\|\cdot\|_2$ norms.

\begin{table}[htbp]
\begin{center}
	\caption{Simulations: Multiple Testing approach}
	\label{tab:sim:het:subsamples}
	\begin{tabular}{c|ccc}
  \hline\hline
	&\multicolumn{3}{c}{Multiple testing based on $\Delta$}\\
	 \hline
 sample size & Bonf. & $\|\cdot\|_\infty$ & $\|\cdot\|_2$ \\
   \hline
   &\multicolumn{3}{c}{Ass.\ \ref{ass4} and Ass.\ \ref{ass5} hold ($\delta=0$, $\gamma=0$)} \\
   \hline
  4000 & 0.164 &  0.173 & 0.247 \\
  12000 &  0.154 &  0.158 & 0.245  \\
 \hline
  &\multicolumn{3}{c}{Ass.\ \ref{ass4} and Ass.\ \ref{ass5} violated ($\delta=2$, $\gamma=0.1$)} \\
   \hline
   4000  & 0.302 & 0.308 & 0.408 \\
   12000  & 0.824 & 0.829 & 0.869   \\
   \hline
	  &\multicolumn{3}{c}{Ass.\ \ref{ass4} violated and Ass.\ \ref{ass5} holds ($\delta=2$, $\gamma=0$)} \\
   \hline
   4000  & 0.365 & 0.377 & 0.458  \\
   12000  & 0.867 & 0.874 & 0.915  \\
	 \hline
		  &\multicolumn{3}{c}{Ass.\ \ref{ass4} holds and Ass.\ \ref{ass5} violated ($\delta=0$, $\gamma=0.1$)} \\
   \hline
   4000  & 1.000 & 1.000 & 1.000  \\
   12000  & 1.000 & 1.000 & 1.000  \\
   \hline
\end{tabular}
\end{center}
\par
{\scriptsize Notes: column `Bonf.' corresponds to rejection rate using Bonferroni correction for multiple testing. The columns `$\|\cdot\|_\infty$' and `$\|\cdot\|_2$' correspond to rejection rate using the multiplier bootstrap procedure described in (12) based on the norms $\|\cdot\|_\infty$ and $\|\cdot\|_2$, respectively. The level of statistical significance is 0.1 (or 10\%).}
\end{table}

Table \ref{tab:sim:het:subsamples} reports the results for sample sizes of $n=4000$ and $n=12000$. Under the satisfaction of both Assumptions \ref{ass4} and \ref{ass5} in the upper panel, the rejection rates of the Bonferroni correction and the multiplier bootstrap using the $\|\cdot\|_\infty$ norm  non-negligibly exceed the nominal size of 10\%, while the size distortion is particularly severe for the $\|\cdot\|_2$ norm. When considering violations of one or both assumptions, the test has only limited power under a sample size of $n=4000$. This was the motivation for also considering a substantially larger sample size of $n=12000$, in which power is decent, in particular when considering the multiplier bootstrap with the $\|\cdot\|_2$ norm. Concerning the power issues, it is worth noting that testing within a subset relies on approximately $333$ or $1000$ observations under the smaller or larger sample size, respectively,  since we consider $6$ subsets in the second half of the sample for the DR approach.


\section{Empirical application}\label{application}

This section provides an empirical application to US census data from \citet{Angrist+98}, who used the sex ratio of a mother's first two children as instrument $Z$ for estimating the effect of (higher) fertility, defined as a treatment dummy $D$ for having at least three children, on mothers' labor supply outcomes. The latter include mothers' weeks in employment per year, which we consider as outcome $Y$. The intuition for this IV strategy is that if parents have a preference for mixed sex children, then getting two children of the same sex, which is arguably randomly assigned by nature, increases the chance of having a third child. This implies a positive first stage effect of $Z$, defined as a dummy for having two kids of the same sex, on $D$. 

However, \citet{RoWo00} argue that having mixed sex siblings may violate IV validity by directly affecting both the marginal utility of leisure and child rearing costs and, thus, labor supply.\footnote{Furthermore, for a data set from rural India, they also provide empirical evidence that expenses for clothing of the third born are significantly lower if the older siblings are of the same sex, which may also affect labor supply decisions and thus, violate the exclusion restriction. Even though it is not clear to which extent this issue carries over to the US data of \citet{Angrist+98}, concerns about IV validity remain.} In addition, a further IV assumption required for the identification of the LATE might be challenged, namely monotonicity, which states that no mother (or family) in the population prefers two children of the same sex over mixed sex children. \citet{Lee07f} for instance finds that South Korean parents with one son and one daughter are more likely to continue childbearing than parents with two sons and it might be doubted that such cases can be fully excluded in the US data of \citet{Angrist+98}. Here, we do not \textit{a priori} impose the IV assumptions required for LATE identification, but instead use the sibling sex ratio instrument to jointly test IV validity and the selection-on-observables assumption. 

To this end, we consider a subsample of the 1980 wave in \citet{Angrist+98} from the US Census Public Use Micro Samples, namely married white couples with mother's education amounting to 12 years, all in all 143,410 observations. While this sample restriction already controls for ethnicity and mother's education, we in addition consider mother's age, mother's age at first birth, father's age, and father's income as covariates $X$. We estimate $\Delta$ and $\theta$ based on lasso regression as ML, using the `causalweight' package by \citet{BodoryHuber2018}. To make the lasso regressions more flexible, we also include interaction terms between the variables in the conditioning set as well as squared and cubic terms of the age and income variables. Table \ref{tab:res1} presents the estimates (`est') of $\Delta$ and $\theta$, respectively, along with the standard errors (`se') and p-values (`pval'). The violation of our testable implication is statistically significant
at the 5\% level based on DR estimation of $\Delta$ (`lasso-based DR') and even considerably more significant (with a p-value close to zero) when applying the squared difference-based test based on $\theta$ (`lasso-based SD').

\begin{table}[htbp]
\begin{center}
	\caption{DR and squared difference-based tests}
	\label{tab:res1}
	\begin{tabular}{c|ccc}
  \hline\hline
 & est & se & pval \\
  \hline
  lasso-based DR  & 0.22 & 0.11 & 0.05\\
	lasso-based SD  & 0.5 & 0.02 & 0.00\\
   \hline
\end{tabular}
\end{center}
\par
{\scriptsize Notes: columns `est', `se', and `pval' provide the estimates, the standard errors and the p-values for $\Delta$ and $\theta$, respectively.}
\end{table}

Finally, we aim at estimating the violations based on DR within subsets of the data which are defined as a function of elements in $(D,X)$ that importantly predict violations of our testable implication, as discussed in Section \ref{DRtest}. We randomly split our sample into two halves and proceed analogously as in the simulations in Section \ref{simulation}. In the first subsample, we estimate the score functions based on cross-fitting and use the random forest of the `grf' package to estimate the nuisance parameters, We apply another random forest for determining the variable importance of elements in $(D,X)$ in predicting the estimated scores based on the `randomForest' package. The results suggest that father's income is by far the most important predictor of heterogeneity in violations $\Delta(D,X)$. Hence, we split the second subsample into subsets based on the median of father's income and estimate the violation $\theta_m$ based on the DR score function in equation \eqref{DR_sim} in either subset and test statistical significance separately and jointly across subsets.

\begin{table}[htbp]
\begin{center}
	\caption{DR tests within subsets}
	\label{tab:res2}
	\begin{tabular}{c|c|ccc|ccc}
  \hline\hline
splitting variable &  subset & est & se & pval & bonf & boot ($l_\infty$) & boot ($l_2$)\\
  \hline
father's income & in lower half & 0.076 & 0.162 & 0.641 & & &\\
&in upper half & 0.328 & 0.152 & 0.030 & 0.0606 & 0.0588 & 0.0825\\
   \hline
\end{tabular}
\end{center}
\par
{\scriptsize Notes: columns `est', `se', and `pval' provide the estimates for $\Delta$ and the respective standard errors and p-values within subsets defined upon the quantiles of the variables in the first column. The three last columns `bonf', `boot ($l_\infty$)', and `boot ($l_2$)' yield the p-value for the joint significance across all subsets using Bonferroni correction and multiplier bootstrap based on the $l_\infty$-norm and $l_2$-norm, respectively.}
\end{table}

Table \ref{tab:res2} reports the results of our DR approach across subsets. It provides the estimated violation (`est') separately for observations in the lower and upper halves of the distribution of father's income. The estimate in the upper half (0.328) is statistically significantly different from zero at the 5\% level (see `pval'), while that in the lower one is close to zero and insignificant. When testing the null hypothesis in both subsets jointly, the p-values amounts to roughly 6\% when using the Bonferroni correction or the multiplier bootstrap based on the $l_\infty$-norm. The p-value of the test using multiplier bootstrap based on the $l_2$-norm is slightly higher (8.25\%).
By and large, the results of our statistical tests presented in Tables \ref{tab:res1} and \ref{tab:res2}
point to a violation of the testable implication in equation \eqref{mainresult2}, such that identification might fail. Given the limited set of covariates in our empirical illustration, our findings are most likely driven by a violation of the selection-on-observables assumption, even though IV validity cannot be taken for granted either for the reasons discussed at the beginning of this section.


\section{Conclusion}\label{co}
In this paper, we demonstrated the existence of a testable condition for the identification of treatment effects, given that a set of covariates to be controlled for and a \textit{suspected} instrument are available in the data and specific assumptions about the causal relations of the observed variables hold. The testable condition corresponds to the conditional independence of the \textit{suspected} instrument and the outcome given the treatment and the covariates. It at the same time implies that the \textit{suspected} instrument is valid (i.e.\ satisfies the exclusion restriction and is as good as random conditional on the covariates) and that the treatment is as good as random conditional on the covariates such that the treatment is identified given the covariates. 

We proposed tests of this conditional independence based on doubly robust estimators aimed at either detecting average violations in the total sample or violations in subsamples in a data-driven way as a function of observed covariates. We also suggested a further, global test accounting for covariate-related heterogeneities in violations which is based on the average squared violation in the sample. We derived the asymptotic distribution of newly suggested tests and investigated the finite sample behavior of all approaches in a simulation study. Further, we applied our tests to the evaluation of the impact of fertility on female labor supply when using the sibling sex ratio of the first two children as supposed instrument. Our results pointed to a violation of our testable implication under the moderate set of socio-economic covariates considered.

\pagebreak

\begin{appendix}

\section{Appendix A: Proofs of Results}\label{proofs}
\subsection{Proof of Theorem \ref{theorem1}}\label{proofth1}

We subsequently provide a proof for Theorem \ref{theorem1}. The latter states that
conditional on Assumptions \ref{ass0} and \ref{ass3}, $Y {\perp\!\!\!\perp}  Z | D=d, X$ is a necessary and sufficient condition for the joint satisfaction of $Y(d) {\perp\!\!\!\perp}  D |  X$ and  $Y(d) {\perp\!\!\!\perp}  Z |  X$. That is, Assumptions \ref{ass1} and \ref{ass2} hold when considering potential outcomes $Y(d)$ matching the factual treatment assignment $D=d$.

We show this result based on a proof by contradiction. To this end, let us assume that in addition to Assumptions \ref{ass0} and \ref{ass3}, also Assumptions \ref{ass1} and \ref{ass2} hold, i.e.\ $Y(d) {\perp\!\!\!\perp}  D |  X$ and $Y(d) {\perp\!\!\!\perp}  Z |  X$. Furthermore, assume a violation of $Y {\perp\!\!\!\perp}  Z | D=d, X$, which by the observational rule  equals  $Y(d) {\perp\!\!\!\perp}  Z | D=d, X$, because $Y=Y(d)$ if $D=d$. The latter violation necessarily implies that conditional on $X$, $D$ is a collider between $Z$ and $Y(d)$ in the sense of \citet{Pearl00}, meaning that controlling for $D$ introduces statistical dependence between $Z$ and $Y(d)$. To see this, first note that by Assumptions \ref{ass2} and \ref{ass3}, $Z$ is independent of $Y(d)$ and associated with $D$, respectively, conditional on $X$. Given the causal framework and faithfulness imposed by Assumption \ref{ass0}, the conditional dependence between $Z$ and $D$ may only be due to a causal effect of $Z$ on $D$ and/or to unobserved confounders affecting both $Z$ and $D$. If conditioning on $D$ introduces statistical dependence between $Z$ and $Y(d)$ (collider bias), then there must necessarily exist a confounder affecting both $D$ and $Y(d)$ conditional on $X$. This follows from a combination of the so-called d-separation criterion of \citet{pearl1988probabilistic}, which implies that $D$ is a collider if (i) a confounder affects both $D$ and $Y(d)$ or (ii) $Y(d)$ affects $D$, and Assumption \ref{ass0}, which rules out case (ii) that $Y(d)$ affects $D$ (reverse causality). However, confounding of $D$ and $Y(d)$ conditional on $X$ violates (and thus, contradicts) Assumption \ref{ass1}, namely $Y(d) {\perp\!\!\!\perp}  D |  X$. This in turn implies that if Assumption \ref{ass1} holds such that $D$ is not a collider and controlling on $D$ does not introduce a dependence of $Y(d)$ and $Z$ given $X$, then  $Y(d) {\not\perp\!\!\!\perp}  Z | D, X$ must be due to statistical dependence between $Z$ and $Y(d)$ conditional $X$. The latter, however, violates (and thus, contradicts) Assumption \ref{ass2}.

We have demonstrated that the joint satisfaction of Assumptions \ref{ass0}, \ref{ass3}, \ref{ass1}, and \ref{ass2} necessarily implies $Y(d) {\perp\!\!\!\perp}  Z | D=d, X$. Next, we also show the converse, namely that conditional on Assumptions \ref{ass0} and \ref{ass3}, the satisfaction of $Y(d) {\perp\!\!\!\perp}  Z | D=d, X$ necessarily implies the joint satisfaction of Assumptions \ref{ass1} and \ref{ass2}. In our proof by contradiction, we assume that $Y(d) {\perp\!\!\!\perp}  Z | D=d, X$ holds while Assumptions \ref{ass1} and/or \ref{ass2} are violated. The latter violations imply that conditional on $X$, there exist confounders of $D$ and $Y(d)$ and/or statistical dependence between $Z$ and $Y(d)$ (due to confounders or a violation of the exclusion restriction). Under the causal faithfulness assumption imposed in Assumption \ref{ass0} (which rules out the possibility that multiple violations exactly cancel each other out), any such violation implies $Y(d) {\not\perp\!\!\!\perp}  Z | D=d, X$, thereby leading to a contradiction. Therefore, it follows that
\begin{eqnarray}\label{nesandsuf}
Y(d) {\perp\!\!\!\perp}  D |  X,\quad Y(d) {\perp\!\!\!\perp}  Z |  X \iff  Y {\perp\!\!\!\perp}  Z | D=d, X \quad\forall d \in \mathcal{D},
\end{eqnarray}
such that $Y {\perp\!\!\!\perp}  Z | D=d, X$ is necessary and sufficient for Assumption \ref{ass1} and \ref{ass2}, conditional on Assumptions \ref{ass0} and \ref{ass3}.


\subsection{Proof of Theorem \ref{theorem2}}\label{proofth2}

We subsequently provide a proof for Theorem \ref{theorem2}. The latter  states that conditional on Assumptions \ref{ass0} and Assumption \ref{ass6},  $E[Y|Z,D=d,X]=E[Y|D=d,X]$ is a necessary and sufficient condition for the joint satisfaction of  $E[Y(d)| D=d, X]=E[Y(d)|  X]$ and $E[Y(d)| Z, X]=E[Y(d)|  X]$. That is, Assumptions 6 and \ref{ass5} hold when considering potential outcomes $Y(d)$ that match the factual treatment assignment $D=d$.

We show this result based on a proof by contradiction. To this end, let us assume that in addition to Assumptions \ref{ass0} and \ref{ass6},  also Assumptions \ref{ass4} and \ref{ass5} hold. Furthermore, assume a violation of $E[Y|Z,D=d,X]\neq E[Y|D=d,X]$, which corresponds to $E[Y(d)|Z,D=d,X]\neq E[Y(d)|D=d,X]$ by the observational rule, because $Y=Y(d)$ if $D=d$. The latter inequality necessarily implies that conditional on $X$, $D$ is a collider between $Z$ and $Y(d)$ in the sense of \citet{Pearl00}, meaning that controlling for $D$ in addition to $X$ introduces a non-zero correlation between $Z$ and $Y(d)$. To see this, first note that by Assumptions \ref{ass5} and \ref{ass6}, $Z$ is mean independent of $Y(d)$, but not of $D$ conditional on $X$. Given the causal framework and faithfulness imposed by Assumption \ref{ass0}, the conditional dependence between $Z$ and $D$ may only be due to a non-zero average causal effect of $Z$ on $D$ and/or to unobserved confounders affecting both $Z$ and $D$. If conditioning on $D$ introduces mean dependence between $Z$ and $Y(d)$, then there must necessarily exist a confounder affecting both $D$ and $Y(d)$ conditional on $X$. This follows from a combination of the so-called d-separation criterion of \citet{pearl1988probabilistic}, which implies that $D$ is a collider if (i) a confounder affects both $D$ and $Y(d)$ or (ii) $Y(d)$ has a non-zero average effect on $D$, and Assumption \ref{ass0}, which rules out case (ii) that $Y(d)$ has any effect on $D$ (reverse causality). However, confounding of $D$ and $Y(d)$ conditional on $X$ violates (and thus, contradicts) Assumption 6. This in turn implies that if Assumption \ref{ass4} holds such that $D$ is not a collider and controlling on $D$ does not introduce a correlation between $Y(d)$ and $Z$ conditional on $X$, then $E[Y(d)|Z,D=d,X]\neq E[Y(d)|D=d,X]$ must be due to a correlation of $Z$ and $Y(d)$ conditional $X$. The latter, however, violates (and thus, contradicts) Assumption \ref{ass5}.

We have demonstrated that the joint satisfaction of Assumptions \ref{ass0}, \ref{ass6}, \ref{ass4}, and \ref{ass5} necessarily implies $E[Y(d)|Z,D=d,X]= E[Y(d)|D=d,X]$. Next, we also show the converse, namely that conditional on Assumptions \ref{ass0} and \ref{ass6}, $E[Y(d)|Z,D=d,X]=E[Y(d)|D=d,X]$ necessarily implies the joint satisfaction of Assumptions 6 and \ref{ass5}. In our proof by contradiction, we assume that $E[Y(d)|Z,D=d,X]= E[Y(d)|D=d,X]$ holds while Assumptions \ref{ass4} and/or \ref{ass5} are violated. The latter violations imply that conditional on $X$, there exist confounders of $D$ and $Y(d)$ and/or a correlation of $Z$ and $Y(d)$ (due to confounders or a violation of the exclusion restriction). Under causal faithfulness (imposed in  Assumption \ref{ass0}), any of these violations imply $E[Y(d)|Z,D=d,X]\neq E[Y(d)|D=d,X]$, thereby leading to a contradiction. Therefore, it follows that
\begin{eqnarray}\label{nesandsuf2}
&&E[Y(d)| D, X]=E[Y(d)|  X],\quad E[Y(d)| Z, X]=E[Y(d)|  X]\\ &\iff&  E[Y|Z,D=d,X]= E[Y|D=d,X] \quad\forall d \in \mathcal{D},\notag
\end{eqnarray}
such that $E[Y|Z,D=d,X]= E[Y|D=d,X]$ is necessary and sufficient for Assumption \ref{ass4} and \ref{ass5}, given Assumptions \ref{ass0} and \ref{ass6}.


\subsection{Proof of Theorem \ref{theoremDR}}\label{proofDR}

To prove Theorem \ref{theoremDR}, we adapt the proof of Theorem 5.1 in \cite{doubleML}. All bounds in the proof hold uniformly over $P\in\mathcal{P}$ but we omit this qualifier for brevity. We use $C$ to denote a strictly positive constant that is independent of $n$ and $P\in\mathcal{P}$. The value of $C$ may change at each appearance.

\begin{proof}
First, we observe that the score in \eqref{DR_sim} is linear
\begin{align*}
\phi_m(W,\theta,\eta)=\phi_m^a(W,\eta)\theta+\phi_m^b(W,\eta)
\end{align*}
with $\phi_m^a(W,\eta)=-1$ and
\begin{align*}
\phi_m^b(W,\eta)&=\Bigg(\mu(1,D,X)-\mu(0,D,X)\\
&+\frac{(Y-\mu(1,D,X))\cdot Z}{p(D,X)}
-\frac{(Y-\mu(0,D,X))\cdot (1-Z)}{1-p(D,X)}\Bigg)I\{D,X \in L_m \}.
\end{align*}
The score fulfills the moment condition, e.g.
\begin{align*}
&\quad\mathbb{E}[\phi_m(W,\theta_0,\eta_0)]\\
&=\mathbb{E}\left[\left(\frac{(Y-\mu(1,D,X))\cdot Z}{p(D,X)}
-\frac{(Y-\mu(0,D,X))\cdot (1-Z)}{1-p(D,X)}\right)I\{ D,X \in L_m \}\right]\\
&=0
\end{align*}
for each $m=1,\dots,M$ by construction of the score since $\mu(Z,D,X)=\mathbb{E}[Y|Z,D,X]$. Next, we verify Neyman orthogonality. For all $\eta=(\mu,p)$, it holds
\begin{align*}
&\quad\partial_r\mathbb{E}[\phi_m(W,\theta_0,\eta_0 + r(\eta-\eta_0)]\big|_{r=0}\\
&=\mathbb{E}\left[\partial_r\phi_m(W,\theta_0,\eta_0 + r(\eta-\eta_0))\big|_{r=0}\right]\\
&=\mathbb{E}\left[(\mu(1,D,X)-\mu_0(1,D,X))I\{ D,X \in L_m \}\right]-\mathbb{E}\big[(\mu(0,D,X)\\
&\quad-\mu_0(0,D,X))I\{D,X \in L_m \}\big]
-\mathbb{E}\left[\frac{Z(\mu(1,D,X)-\mu_0(1,D,X))}{p_0(D,X)}I\{D,X \in L_m \}\right]\\
&\quad+\mathbb{E}\left[\frac{(1-Z)(\mu(0,D,X)-\mu_0(0,D,X))}{1-p_0(D,X)}I\{D,X \in L_m \}\right]\\
&\quad-\mathbb{E}\left[\frac{Z(Y-\mu_0(1,D,X))(p(D,X)-p_0(D,X))}{p_0(D,X)^2}I\{ D,X \in L_m \}\right]\\
&\quad-\mathbb{E}\left[\frac{(1-Z)(Y-\mu_0(0,D,X))(p(D,X)-p_0(D,X))}{(1-p_0(D,X))^2}I\{ D,X \in L_m \}\right]=0
\end{align*}
since $\mathbb{E}[Z|D,X]=p_0(D,X)$, $\mathbb{E}[Z(Y-\mu_0(1,D,X))|D,X]=0$ and 
$$\mathbb{E}[(1-Z)(Y-\mu_0(0,D,X))|D,X]=0$$ 
by definition.
Also, the identification condition holds, namely,
$$J_0:=\mathbb{E}[\phi_2^a(W,\eta)]=-1.$$
This gives us Assumption 3.1 in \cite{doubleML}. To prove Assumption 3.2 d) in \cite{doubleML}, we observe that 
$$\Sigma^2:=\mathbb{E}[\phi_M(W,\theta_0,\eta_0)\phi_M(W,\theta_0,\eta_0)^T]$$
with diagonal elements
\begin{align*}
&\quad\mathbb{E}[\phi_m(W,\theta_0,\eta_0)^2]\\
&=\mathbb{E}[\mathbb{E}[\phi_m(W,\theta_0,\eta_0)^2|X,D]]\\
&=\mathbb{E}\Big[\mathbb{E}\Big[\Big(\Big((\mu(1,D,X)-\mu(0,D,X)\big)I\{ D,X \in L_m \}-\theta_m\Big)\\
&\quad+\Bigg(\frac{(Y-\mu(1,D,X))\cdot Z}{p(D,X)}
-\frac{(Y-\mu(0,D,X))\cdot (1-Z)}{1-p(D,X)}\Bigg)I\{ D,X \in L_m \}\Big)^2\Big]\Big|X,D\Big]\\
&=\mathbb{E}\Big[\mathbb{E}\Big[\Big(\Big(\mu(1,D,X)-\mu(0,D,X)\big)I\{ D,X \in L_m \}-\theta_m\Big)^2\\
&\quad+\Bigg(\frac{(Y-\mu(1,D,X))\cdot Z}{p(D,X)}
-\frac{(Y-\mu(0,D,X))\cdot (1-Z)}{1-p(D,X)}\Bigg)^2I\{ D,X \in L_m \}\Big]\Big|X,D\Big]
\end{align*}
with the same arguments as above. Thus, due to Assumption \ref{assDR},
\begin{align*}
&\quad\mathbb{E}[\phi_m(W,\theta_0,\eta_0)^2]\\
&\ge\mathbb{E}\Bigg[\Bigg(\frac{(Y-\mu(1,D,X))\cdot Z}{p(D,X)}
-\frac{(Y-\mu(0,D,X))\cdot (1-Z)}{1-p(D,X)}\Bigg)^2I\{ D,X \in L_m \}\Bigg]\\
&=\mathbb{E}\Bigg[\Bigg(\frac{(Y-\mu(1,D,X))^2\cdot Z^2}{p(D,X)}
+\frac{(Y-\mu(0,D,X))^2\cdot (1-Z)^2}{1-p(D,X)}\Bigg)I\{ D,X \in L_m \}\Bigg]\\
&\ge\frac{1}{(1-\epsilon)^2}\mathbb{E}\Bigg[\Bigg((Y-\mu(1,D,X))^2\cdot Z^2
+(Y-\mu(0,D,X))^2\cdot (1-Z)^2\Bigg)I\{ D,X \in L_m \}\Bigg]\\
&=\frac{1}{(1-\epsilon)^2}\mathbb{E}\big[U^2I\{ D,X \in L_m \}\big]\\
&\ge\frac{c}{(1-\epsilon)^2}.
\end{align*}
Further, for $m\neq n$, it holds
\begin{align*}
\mathbb{E}[\phi_m(W,\theta_0,\eta_0)\phi_n(W,\theta_0,\eta_0)]=0
\end{align*}
since $I\{ D_i,X_i \in L_m \}I\{ D_i,X_i \in L_n \}=0$ almost surely and $\theta_n=\theta_m=0$ under $H_0$. Thus, all eigenvalues of $\Sigma^2$ are bounded from below under $H_0$.
Next, we show Assumption 3.2 a)-c) to complete the proof. We define the following nuisance realization set $\mathcal{T}_n$ as the set of all P-square-integrable functions $\eta=(\eta_1,\eta_2)=(\mu,p)$ such that
\begin{align*}
\max(\|(\eta_{0,1}-\eta_1)\|_{P,2},\|(\eta_{0,2}-\eta_2)\|_{P,2})&\le C\\
\max(\|\eta_{0,1}-\eta_1\|_{P,q},\|\eta_{0,2}-\eta_2\|_{P,q})&\le \delta_N\\
\|\eta_2-1/2\|_{P,\infty}&\le 1/2-\epsilon\\
\|\eta_{0,1}-\eta_1\|_{P,2}\times \|\eta_{0,2}-\eta_2\|_{P,2}&\le\delta_NN^{-1/2}
\end{align*}
for $\delta_N=o(1)$ and a constant $q>2$. Note that Assumption 3.2(a) holds by construction of the set $\mathcal{T}_N$ and Assumption \ref{assDR}. Next, we verify Assumption 3.2(b). First, we note that
\begin{align*}
\left(\|\mu_0(0,D,X)\|_{P,q}\vee \|\mu_0(0,D,X)\|_{P,q}\right)\le \|\mu_0(Z,D,X)\|_{P,q}/\epsilon^{1/q}\le\|Y\|_{P,q}/\epsilon^{1/q}\le C/\epsilon^{1/q},
\end{align*}
due to Assumption \ref{assDR} i) where we use that $P(Z=1|D,X)=p_0(D,X)\ge \epsilon$ and $P(Z=0|D,X)=1-p_0(D,X)\ge \epsilon$. Similarly, for any $\mu\in\mathcal{T}_n$, it holds
\begin{align*}
\|\mu(1,D,X)-\mu_0(1,D,X)\|_{P,q}\le C/\epsilon^{1/q},\quad \|\mu(0,D,X)-\mu_0(0,D,X)\|_{P,q}\le C/\epsilon^{1/q}
\end{align*}
and
\begin{align*}
|\theta_m|\le |E[(\mu_0(1,D,X)-\mu_0(0,D,X))]|\le 2C/\epsilon^{1/q}.
\end{align*}
Therefore, for any $\eta\in\mathcal{T}_n$, it holds
\begin{align*}
&\quad\mathbb{E}[\|\phi_M(W,\theta_0,\eta_0)\|^q]^{1/q}\\
&\le \sqrt{M} \mathbb{E}\left[\left(\sup_{m=1,\dots,M}\phi_m(W,\theta_0,\eta_0)\right)^q\right]^{1/q}\\
&\le \sqrt{M}\left((1+\epsilon^{-1})\left(\|\mu(1,D,X)\|_{P,q}+ \|\mu(0,D,X)\|_{P,q}\right)+\|Y\|_{P,q}/\epsilon+2C/\epsilon^{1/q}\right)\\
&\le \sqrt{M}\left(4(1+\epsilon^{-1})/\epsilon^{1/q}+2C/\epsilon+2C/\epsilon^{1/q}\right).
\end{align*}
Also, we have
\begin{align*}
\sup_{\eta\in\mathcal{T}_N}\mathbb{E}[|\phi_m^a(W,\theta_0,\eta)|^q]^{1/q}&=1.
\end{align*}
Finally, we verify Assumption 3.2 c). It holds
\begin{align*}
\sup_{\eta\in\mathcal{T}_N}|\mathbb{E}[\phi_m^a(W,\theta_0,\eta)-\phi_2^a(W,\theta_0,\eta_0)]|=|(-1)-(-1)|=0.
\end{align*}
Analogously to the verification of Assumption 3.2 b) following the proof of Theorem 5.1 in \cite{doubleML}, we conclude
\begin{align*}
&\quad\sup_{\eta\in\mathcal{T}_N}\mathbb{E}[\|\phi_M(W,\theta_0,\eta)-\phi_M(W,\theta_0,\eta_0)\|^2]^{1/2}\\
&\le\sqrt{M} \sup_{\eta\in\mathcal{T}_N}\mathbb{E}\left[\sup\limits_{m=1,\dots,M}\left(\phi_m(W,\theta_0,\eta)-\phi_m(W,\theta_0,\eta_0)\right)^2\right]^{1/2}\\
&\le \sqrt{M}\left(I_1+I_{II}+I_{III}\right)
\end{align*}
with
\begin{align*}
I_1:=\|\mu(1,D,X)-\mu_0(1,D,X)\|_{P,q}+\|\mu(0,D,X)-\mu_0(0,D,X)\|_{P,2}\le 2\delta_n/\epsilon^{1/2},
\end{align*}
\begin{align*}
I_{II}:=\left\|\frac{(Y-\mu(1,D,X))\cdot Z}{p(D,X)}-\frac{(Y-\mu_0(1,D,X))\cdot Z}{p_0(D,X)}\right\|_{P,2}\le \epsilon^{-2}\left(\epsilon^{-1/2}+\sqrt{C}\right)2\delta_n
\end{align*}
and
\begin{align*}
I_{III}&:=\left\|\frac{(Y-\mu(0,D,X))\cdot (1-Z)}{1-p(D,X)}-\frac{(Y-\mu_0(0,D,X))\cdot (1-Z)}{1-p_0(D,X)}\right\|_{P,2}\\
&\le \epsilon^{-2}\left(\epsilon^{-1/2}+\sqrt{C}\right)2\delta_n.
\end{align*}
Finally,
\begin{align*}
&\quad\sup_{\eta\in\mathcal{T}_N, r\in (0,1)}\|\partial_r^2\mathbb{E}[\phi_M(W,\theta_0,\eta_0 + r(\eta-\eta_0)]\|\\
&\le\sqrt{M}\sup_{\eta\in\mathcal{T}_N, r\in (0,1), m=1,\dots,M}\mathbb{E}\left[\partial_r^2\phi_m(W,\theta_0,\eta_0 + r(\eta-\eta_0))\right]\\
&\le\delta_NN^{-1/2}
\end{align*}
completes the proof.
\end{proof}

\subsection{Proof of Theorem \ref{theorem3}}\label{proofth3}

To prove Theorem \ref{theorem3}, we apply Theorem 3.1 in \cite{doubleML} (analog to Theorem 5.1 and 5.2). All bounds in the proof hold uniformly over $P\in\mathcal{P}$ but we omit this qualifier for brevity. We use $C$ to denote a strictly positive constant that is independent of $n$ and $P\in\mathcal{P}$. The value of $C$ may change at each appearance.

\begin{proof}
First, we observe that the score in \eqref{score2} is linear
\begin{align*}
\phi_2(W,\theta,\eta)=\phi_2^a(W,\eta)\theta+\phi_2^b(W,\eta)
\end{align*}
with $\phi_2^a(W,\eta)=-1$ and $\phi_2^b(W,\eta)=(\eta_1(W)-\eta_2(W))^2+\zeta$. The score fulfills the moment condition, e.g.
\begin{align*}
\mathbb{E}[\phi_2(W,\theta_0,\eta_0)]=\mathbb{E}[(\mu(1,X,D)-\mu(0,X,D))^2-\mathbb{E}[(\mu(1,X,D)-\mu(0,X,D))^2]+\zeta]=0
\end{align*}
by construction of the score. Under $H_0$, it holds
\begin{align*}
&\partial_r\mathbb{E}[\phi_2(W,\theta_0,\eta_0 + r(\eta-\eta_0)]\big|_{r=0}\\
&=\mathbb{E}\left[\partial_r\phi_2(W,\theta_0,\eta_0 + r(\eta-\eta_0))\big|_{r=0}\right]\\
&=\mathbb{E}\left[\partial_r\left(\eta_{0}(1,X,D)+r(\eta-\eta_{0})(1,X,D)-(\eta_{0}(0,X,D)+r(\eta-\eta_{0})(0,X,D))\right)^2\big|_{r=0}\right]\\
&=\mathbb{E}[2\underbrace{(\eta_{0}(1,X,D)-\eta_{0}(0,X,D))}_{=0 \text{\ a.s.}}((\eta-\eta_{0})(1,X,D)-(\eta-\eta_{0})(0,X,D))]=0
\end{align*}
and therefore the score fulfills the Neyman orthogonality condition. Also, the identification condition holds, namely,
$J_0:=\mathbb{E}[\phi_2^a(W,\eta)]=-1$. This gives us Assumption 3.1 in \cite{doubleML}. Next, we show Assumption 3.2 to complete the proof. We define the following nuisance realization set $\mathcal{T}_n$ as the set of all P-square-integrable functions $\eta$ such that
\begin{align*}
\|\eta_{0}-\eta\|_{P,2q}&\le C,\\
\|\eta_{0}-\eta\|_{P,4}&\le \delta_N,\\
\|\eta_{0}-\eta\|_{P,2}&\le\delta_N^{1/2}N^{-1/4}
\end{align*}
for $\delta_N=o(1)$ and a constant $q>2$. Note that Assumption 3.2(a) holds by construction of the set $\mathcal{T}_N$ and Assumption \ref{ass7}. Next, we verify Assumption 3.2(b). For $q>2$, we have
\begin{align*}
\sup_{\eta\in\mathcal{T}_N}\mathbb{E}[|\phi_2(W,\theta_0,\eta)|^q]^{1/q}&=\sup_{\eta\in\mathcal{T}_N}\mathbb{E}\left[\left((\eta(1,X,D)-\eta(0,X,D))^2-\theta_0+\zeta\right)^q\right]^{1/q}\\
&\le \sup_{\eta\in\mathcal{T}_N}\|(\eta(1,X,D)-\eta(0,X,D))^2\|_{P,q}+|\theta_0|+\|\zeta\|_{P,q}\\
&\lesssim \sup_{\eta\in\mathcal{T}_N}\|(\eta(1,X,D)-\eta(0,X,D))^2\|_{P,q}+C
\end{align*}
due to $\|\zeta\|_{P,q}<C$ with
\begin{align*}
&\quad\sup_{\eta\in\mathcal{T}_N}\|(\eta(1,X,D)-\eta(0,X,D))^2\|_{P,q}\\
&=\sup_{\eta\in\mathcal{T}_N}\|\eta(1,X,D)-\eta(0,X,D)\|_{P,2q}^{2}\\
&\le\sup_{\eta\in\mathcal{T}_N}\|(\eta-\eta_{0})(1,X,D)+(\eta_0(1,X,D)-\eta_0(0,X,D))+(\eta_{0}-\eta)(0,X,D)\|_{P,2q}^2\\
&\le \sup_{\eta\in\mathcal{T}_N}\left(\|(\eta-\eta_{0})(1,X,D)\|_{P,2q}+\|(\eta_{0}-\eta)(0,X,D)\|_{P,2q}\right)^2\\
&\le C
\end{align*}
by construction of the nuisance realization set. Also, we have
\begin{align*}
\sup_{\eta\in\mathcal{T}_N}\mathbb{E}[|\phi_2^a(W,\theta_0,\eta)|^q]^{1/q}&=1.
\end{align*}
Now, we verify Assumption 3.2(c). It holds
\begin{align*}
\sup_{\eta\in\mathcal{T}_N}|\mathbb{E}[\phi_2^a(W,\theta_0,\eta)-\phi_2^a(W,\theta_0,\eta_0)]|=|(-1)-(-1)|=0.
\end{align*}
In addition, we have
\begin{align*}
&\quad\sup_{\eta\in\mathcal{T}_N}\mathbb{E}[|\phi_2(W,\theta_0,\eta)-\phi_2(W,\theta_0,\eta_0)|^2]^{1/2}\\
&=\sup_{\eta\in\mathcal{T}_N}\mathbb{E}\left[\left((\eta(1,X,D)-\eta(0,X,D))^2-(\mu(1,X,D)-\mu(0,X,D))^2\right)^2\right]^{1/2}\\
&=\sup_{\eta\in\mathcal{T}_N}\mathbb{E}\bigg[\left((\mu(1,X,D)-\mu(0,X,D))+(\eta(1,X,D)-\eta(0,X,D))\right)^2\\
&\quad\left((\mu(1,X,D)-\mu(0,X,D))-(\eta(1,X,D)-\eta(0,X,D))\right)^2\bigg]^{1/2}\\
&\le\sup_{\eta\in\mathcal{T}_N}\mathbb{E}\left[((\mu(1,X,D)-\mu(0,X,D))+(\eta(1,X,D)-\eta(0,X,D)))^4\right]^{1/4}\\
&\quad\mathbb{E}\left[((\mu(1,X,D)-\mu(0,X,D))-(\eta(1,X,D)-\eta(0,X,D)))^4\right]^{1/4}\\
&\le C\sup_{\eta\in\mathcal{T}_N}\left(\mathbb{E}\left[((\mu(1,X,D)-\eta(1,X,D))+(\eta(0,X,D)-\mu(0,X,D)))^4\right]^{1/4}\right)\\
&\le 2C\sup_{\eta\in\mathcal{T}_N}\|\eta_{0}-\eta\|_{P,4}\lesssim \delta_N.
\end{align*}
Further, for all $r\in (0,1)$, it holds
\begin{align*}
&\quad\sup_{\eta\in\mathcal{T}_N}\partial_r^2\mathbb{E}[\phi_2(W,\theta_0,\eta_0 + r(\eta-\eta_0)]\\
&=\sup_{\eta\in\mathcal{T}_N}\mathbb{E}\left[\partial_r^2\phi_2(W,\theta_0,\eta_0 + r(\eta-\eta_0))\right]\\
&=\sup_{\eta\in\mathcal{T}_N}\mathbb{E}\big[\partial_r^2(\mu(1,X,D)+r(\eta(1,X,D)-\mu(1,X,D))-\\
&\quad(\mu(0,X,D)+r(\eta(0,X,D)-\mu(0,X,D))))^2\big]\\
&=\sup_{\eta\in\mathcal{T}_N}\mathbb{E}\bigg[\partial_r 2(\mu(1,X,D)+r(\eta(1,X,D)-\mu(1,X,D))\\
&\quad-(\mu(0,X,D)+r(\eta(0,X,D)-\mu(0,X,D))))\\
&\quad((\eta(1,X,D)-\mu(1,X,D))-(\eta(0,X,D)-\mu(0,X,D)))\bigg]\\
&=\sup_{\eta\in\mathcal{T}_N}\mathbb{E}\left[2((\eta(1,X,D)-\mu(1,X,D))-(\eta(0,X,D)-\mu(0,X,D)))^2\right]\\
&\le C \sup_{\eta\in\mathcal{T}_N}\|\eta-\eta_0\|_{P,2}^2\le C(\delta_N^{1/2}N^{-1/4})^2=C\delta_NN^{-1/2}.
\end{align*}
Finally, it is easy to show that the variance of the score $\phi_2$ is non-degenerate, namely
\begin{align*}
\mathbb{E}[\phi_2(W,\theta_0,\eta_0)^2]&=\mathbb{E}[(\mu(1,X,D)-\mu(0,X,D))^2-\theta_0+\zeta)^2]\\
&=\mathbb{E}[(\mu(1,X,D)-\mu(0,X,D))^2-\theta_0)^2]+\mathbb{E}[\zeta^2]\\
&\ge \mathbb{E}[\zeta^2]=\sigma_{\zeta}^2
\end{align*}
since the variance $\sigma_{\zeta}^2$ of $\zeta$ is chosen to be bounded from below. This completes the proof.
\end{proof}

\end{appendix}

\bibliographystyle{qe} 
\bibliography{research_new} 

\end{document}